\documentclass[review,12pt]{elsarticle}

\usepackage{bm}  
\usepackage{amssymb}
\usepackage{amsthm}
\usepackage{amsmath}
\usepackage[linesnumbered,ruled,lined]{algorithm2e}
\graphicspath{{figs/}} 
\usepackage{booktabs}
\usepackage{multirow}
\usepackage{hyperref}

\usepackage[outline]{contour}
\newcommand*{\fancy}[1]{{\color{white}\contour{black}{#1}}}

\journal{X.X.X.}

\begin{document}

\begin{frontmatter}



\title{A generalized essentially non-hourglass total Lagrangian SPH solid dynamics}
\author[myfirstaddress]{Dong Wu}
\ead{dong.wu@tum.de}
\author[mysecondaddress]{Xiaojing Tang}
\ead{xiaojing.tang@tum.de}
\author[mythirdaddress]{Shuaihao Zhang}
\ead{szhang07@connect.hku.hk}
\author[myfirstaddress]{Xiangyu Hu\corref{mycorrespondingauthor}}
\cortext[mycorrespondingauthor]{Corresponding author.}
\ead{xiangyu.hu@tum.de}

\address[myfirstaddress]{Chair of Aerodynamics and Fluid Mechanics, Technical University of Munich, 85748 Garching, Germany}
\address[mysecondaddress]{Chair of Space Propulsion and Mobility, Technical University of Munich, 85521 Ottobrunn, Germany}
\address[mythirdaddress]{Department of Civil Engineering, The University of Hong Kong, Pokfulam, Hong Kong SAR, China}

\begin{abstract}
In this paper, we tackle a persistent numerical instability 
within the total Lagrangian smoothed particle hydrodynamics 
(TLSPH) solid dynamics. 
Specifically, we address the hourglass modes 
that may grow and eventually 
deteriorate the reliability of simulation, 
particularly in the scenarios characterized by large deformations.
We propose a generalized essentially non-hourglass formulation 
based on volumetric-deviatoric stress decomposition, 
offering a general solution 
for elasticity, plasticity, anisotropy, and other material models.
Comparing the standard SPH formulation 
with the original non-nested Laplacian operator 
applied in our previous work \cite{wu2023essentially}
to handle the hourglass issues in standard elasticity, 
we introduce a correction for the discretization of shear stress 
that relies on the discrepancy produced by a tracing-back prediction of 
the initial inter-particle direction
from the current deformation gradient.
The present formulation, 
when applied to standard elastic materials, 
is able to recover the original Laplacian operator. 
Due to the dimensionless nature of the correction, 
this formulation handles complex material models in a very straightforward way. 
Furthermore, 
a magnitude limiter is introduced 
to minimize the correction in domains where the discrepancy is less pronounced. 
The present formulation is validated, with a single set of modeling parameters, 
through a series of benchmark cases, 
confirming good stability and accuracy
across elastic, plastic, and anisotropic materials. 
To showcase its potential, 
the formulation is employed to simulate a complex problem involving  
viscous plastic Oobleck material, contacts, and very large deformation.
\end{abstract}

\end{frontmatter}


\section{Introduction}\label{sec:introduction}
Smoothed particle hydrodynamics (SPH), 
a fully Lagrangian mesh-free method, 
has attracted escalating interest in recent decades 
\cite{randles1996smoothed, luo2021particle, zhang2022review, xu2023methodology, khayyer2023preface}.
In SPH, the continuum is represented by particles, 
and the governing equations are discretized 
through particle interactions based on a Gaussian-like kernel function 
\cite{monaghan2005smoothed, liu2010smoothed, monaghan2012smoothed}. 
As numerous fundamental abstractions, 
intrinsically linked to various physical systems, 
can be effectively represented through particle interactions, 
SPH method has succeeded 
in addressing multi-physics problems within a unified computational framework
\cite{zhang2021sphinxsys, sun2021accurate}, 
including fluid-structure interaction 
\cite{antoci2007numerical, han2018sph, liu2019smoothed, pearl2021sph}, 
cardiac electrophysiology \cite{zhang2021integrative, zhang2023multi}, 
laser beam welding \cite{bierwisch2020particle, sollich2022improved}, 
porous media \cite{ma2022five, lian2023effective, feng2024general}, 
and various other domains.
In such unified computational framework,
the total Lagrangian SPH (TLSPH)
formulation \cite{vignjevic2000treatment, zhang2021multi} 
is often used to model solid dynamics. 

However, 
the numerical instability issue of hourglass modes persists in 
TLSPH solid dynamics, arising from vanishing deformation gradient 
as particles move to a nonphysical zigzag pattern, 
i.e., the zero-energy modes 
\cite{dyka1997stress, vignjevic2000treatment, vignjevic2009review}.
Various strategies have been employed to tackle this issue, 
including introducing a stabilization term 
to the potential energy function \cite{beissel1996nodal} 
or an artificial viscosity term based on 
minimizing the Laplacian of the deformation field \cite{vidal2007stabilized}, 
and additional integration or stress points particles 
to represent the stress field \cite{randles2000normalized, vignjevic2009review}.
While these approaches have shown success in certain benchmarks, 
they often rely on empirical, case-dependent parameters 
or increase algorithmic complexities \cite{islam2019stabilized}.  
Alternatively, 
artificial viscosity similar to 
that used in computational fluid dynamics (CFD) \cite{owen2004tensor}
has been proposed to reduce hourglass modes 
during dynamic simulations \cite{lee2016new, zhang2022artificial}. 
However, these viscosity formulations might face validity concerns 
when the velocity field becomes flat. 
An efficient alternative scheme involves introducing artificial force or stress 
to mitigate discrepancy, arising from the zigzag pattern, 
between the local displacement field and linearly predicted displacement
from the deformation gradient \cite{kondo2010suppressing, ganzenmuller2015hourglass}. 
Despite effectiveness, 
the scheme risks over-stiffening the non-linear part of the displacement field
and still requires case-dependent tuning parameters 
for physically meaningful results 
\cite{belytschko1983correction, stainier1994improved, o2021fluid}.

Recently,
we proposed simple and essential non-hourglass formulations 
based on volumetric-deviatoric stress decomposition
to directly calculate the acceleration induced by shear stress 
through the Laplacian of displacement or velocity 
\cite{wu2023essentially, zhang2023essentially}. 
However, these formulations are restricted for standard elastic materials only. 
Thus, the development of non-hourglass SPH formulations tailored 
for more complex materials,
such as anisotropic material \cite{holzapfel2009constitutive} 
where the stress is biased along specific directions
and plastic material 
where a stress return mapping is necessary 
when the stress state exceeds the yield stress \cite{simo2006computational},
becomes crucial.

In this paper, 
based on volumetric-deviatoric stress decomposition,
we present a generalized essentially non-hourglass TLSPH formulation
suitable for a wide range of materials, 
including elastic, plastic, anisotropic, and other properties. 
Comparing the standard SPH formulation 
with the original non-nested Laplacian operator 
applied in our previous work \cite{wu2023essentially}
to handle the hourglass issues in standard elasticity, 
we introduce a correction for the discretization of shear stress 
that relies on the discrepancy produced by a tracing-back prediction of 
the initial inter-particle direction
from the current deformation gradient.
In cases where only standard elasticity is present, 
this specially designed formulation is able to 
recover the original Laplacian operator. 
It handles complex material models in a very straightforward way, 
given that the correction is formulated in a dimensionless form. 
Furthermore, 
a magnitude limiter is employed to limit 
the correction before the discrepancy reaches a predefined threshold. 
With a single set of modeling parameters, 
extensive benchmark cases are considered 
to validate the stability and accuracy of the present formulation 
for elastic, plastic and anisotropic materials. 
A complex problem, involving  
viscous plastic Oobleck material, contacts and very large deformation,
is also simulated to demonstrate the potential of the proposed formulation.

The structure of this paper is as follows. 
Section \ref{sec:governingeq} describes the total Lagrangian 
kinematics and governing equations of solid dynamics. 
A variety of material models applied in this study 
are outlined in Section \ref{sec:ConstitutiveRelation},
and the present formulation are detailed in Section \ref{sec:SPHalgorithm}. 
Numerical examples are presented and discussed in Section \ref{sec:examples}.
In Section \ref{sec:conclusion}, brief concluding remarks are offered. 
To foster future in-depth investigations,
all computational codes utilized in this study \cite{zhang2020sphinxsys, zhang2021sphinxsys}
are publicly available via the SPHinXsys project website 
 at \url{https://www.sphinxsys.org}.

\section{Kinematics and governing equations}\label{sec:governingeq}
In the context of continuum mechanics within the total Lagrangian framework, 
the kinematics and governing equations are formulated 
with respect to the initial, undeformed reference configuration.
The deformation gradient tensor $\mathbb{F}$ is given by
\begin{equation}\label{deformation-tensor}
	\mathbb{F}  =  \nabla^0 \bm{r} = \nabla^0 \bm{u}  + \mathbb{I},
\end{equation}
where $\bm{u} = \bm{r} - \bm{r}^0$ is the displacement
with $\bm{r}^0$ and $\bm{r}$ denoting the initial and 
current positions of a material point, respectively, 
$\nabla^0 \equiv \frac{\partial}{\partial \bm{r}^0}$ 
the gradient operator with respect to the initial configuration 
and $\mathbb{I} $ the identity matrix.

The governing equations in 
 total Lagrangian formulation can be expressed as
\begin{equation}\label{conservation_equation}
	\begin{cases}
		\rho =  J^{-1}\rho^0 \\
		\rho^0 \ddot {\bm{u}} = \nabla^0  \cdot \mathbb{P}^{\operatorname{T}},
	\end{cases}
\end{equation}
where $\rho^0$ and $\rho$ are the initial and current densities, respectively, 
$J = \det(\mathbb{F})$, 
$\ddot {\bm{u}}$ the acceleration,
$\mathbb{P}$ the first Piola-Kirchhoff stress tensor, 
and $\operatorname{T}$ the matrix transposition operator. 
$\mathbb{P}$ can be obtained 
by the Kirchhoff stress $\fancy{$\tau$}$ as
\begin{equation}
	\label{first_Piola_kirchhoff}
	\mathbb{P} = \fancy{$\tau$}\mathbb{F}^{-\operatorname{T}}. 
\end{equation}

\section{Material models}\label{sec:ConstitutiveRelation}
A series of material models, 
covering elastic, (perfect, linear and non-linear hardening, viscous) plastic, 
anisotropic with fiber direction, 
and electrophysiologically induced active stress model, 
are included here for validating
the proposed non-hourglass formulation. 
Note that the Kirchhoff stress $\fancy{$\tau$}$ used in governing Eqs. \eqref{conservation_equation} 
and \eqref{first_Piola_kirchhoff} 
are decomposed into volumetric and deviatoric components for all models.

\subsection{Standard elastic material}
The Kirchhoff stress $\fancy{$\tau$}$ for the standard elastic material 
is derived form the strain energy function \cite{simo2006computational}
\begin{equation}
	\mathfrak{W}_e = \mathfrak{W}_v \left( J \right) + \mathfrak{W}_s \left(\bar  {\fancy{$b$}} \right).
\end{equation}
Here, the volume-preserving left Cauchy-Green deformation gradient tensor 
is denoted by
$\bar  {\fancy{$b$}} = \left| \fancy{$b$} \right|^{ - \frac{1}{d}} \fancy{$b$}$, 
where $\fancy{$b$} = \mathbb{F}\mathbb{F}^{\operatorname{T}}$. 
For neo-Hookean materials, 
the volume-dependent strain energy $\mathfrak{W}_v \left(J\right)$, 
with the bulk modulus $K$, 
is written as
\begin{equation}
	\mathfrak{W}_v \left( J \right) = \frac{1}{2}K\left[ \frac{1}{2}\left( J^2 - 1 \right) - \ln J \right]. 
\end{equation}
The shear-dependent strain energy $\mathfrak{W}_s \left(\bar  {\fancy{$b$}} \right)$
is expressed as \cite{yue2015continuum}
\begin{equation}
	\mathfrak{W}_s \left( \bar{ \fancy{$b$}} \right) = \frac{1}{2} G \left( \operatorname{tr} \left( \bar {\fancy{$b$}} \right) - d \right), 
\end{equation}
where $d$ denotes the dimension, $G$ the shear modulus.
Subsequently, 
the Kirchhoff stress tensor $\fancy{$\tau$}$ is obtained 
through partial differentiation of the strain energy function 
with respect to the deformation gradient $\mathbb{F}$ as
\begin{equation}
	\label{Kirchhoff_stress}
	\fancy{$\tau$} = \frac{\partial \mathfrak{W}_e}{\partial \mathbb{F} } \mathbb{F}^{\operatorname{T}}  
	= \frac{K}{2}\left( J^2 - 1 \right) \mathbb{I} + G \operatorname{dev} \left( \bar {\fancy{$b$}} \right),
\end{equation}
where 
\begin{equation}
	\label{Kirchhoff_stress_part}
	\operatorname{dev} \left( \bar{ \fancy{$b$}} \right)
	= \bar {\fancy{$b$}} -  \frac{1}{d} \operatorname{tr} \left( \bar{ \fancy{$b$}} \right) \mathbb{I} 
	= \left| \fancy{$b$} \right|^{ - \frac{1}{d}} \left[ \fancy{$b$} - \frac{1}{d} \operatorname{tr} \left( \fancy{$b$} \right) \mathbb{I} \right]
\end{equation}
returns the trace-free part of $\bar{ \fancy{$b$}}$, 
i.e., $\operatorname{tr} \left( \operatorname{dev}\left( \bar{ \fancy{$b$}} \right) \right) = 0$. 

\subsection{Plastic material}
Four distinct plastic models are considered in this study, 
encompassing perfect, linear hardening, non-linear hardening, 
and viscous plastic models. 
We apply the classical $ J_2 $ flow theory \cite{mises1913mechanik}
to characterize the stress-strain evolution in plasticity.
According to this theory, 
the deformation gradient tensor $\mathbb{F}$ 
can be decomposed into its elastic volumetric part $\mathbb{F}_e$ 
and plastic deviatoric part $\mathbb{F}_p$ as \cite{simo2006computational}
\begin{equation} \label{eq:deformationtensor_split}
	\mathbb{F} = \mathbb{F}_e \mathbb{F}_p.
\end{equation}
The elastic part of left Cauchy-Green tensor ${\fancy{$b$}}_e $
is thus defined as ${\fancy{$b$}} = \mathbb{F}_e \mathbb{F}_e^{\operatorname{T}}$. 
For plasticity analysis, 
the plastic right Cauchy deformation gradient tensor $\mathbb{C}_p$ 
is introduced as
\begin{equation} \label{eq:plastic-Lagrangian-tensor}
	\mathbb{C}_p = \mathbb{F}_p^{\operatorname{T}} \mathbb{F}_p.
\end{equation}
The relationship between $\fancy{$b$}_e$ and 
$\mathbb{C}_p$ is described as 
\begin{equation} \label{eq:plastic-tensor}
	\fancy{$b$}_e = \mathbb{F}  \mathbb{C}_p^{-1}  \mathbb{F}^{\operatorname{T}}.
\end{equation}

The plastic behavior is governed by the deviatoric component 
of the Kirchhoff stress which is denoted as
$\fancy{$\tau$}_{de} = G \operatorname{dev} \left( \bar {\fancy{$b$}} \right)$.
To incorporate the plastic behavior,
a scalar yield function $f(\fancy{$\tau$}_{de})$ is introduced.
If $f (\fancy{$\tau$}_{de}) > 0$, 
indicating the material undergoes plasticity, 
$\fancy{$\tau$}_{de}$
will be mapped back by a return mapping to the yield surface, 
a boundary that separates elastic and plastic regions, 
as $\fancy{$\tau$}_{de}^e$. 
The detailed algorithm of the plastic model
is presented in \ref{app:plastic_algorithm}.
It should be emphasized that the updated $\fancy{$b$}_e$ 
obtained through the return mapping process
can be substituted into Eq. \eqref{Kirchhoff_stress} 
to calculate the stress $\fancy{$\tau$}$ for plastic materials
by replacing $\fancy{$b$}$.

\subsection{Holzapfel-Odgen material}
The Holzapfel-Odgen model \cite{holzapfel2009constitutive}
considers the anisotropic nature of the muscle, 
such as myocardium. 
Following Ref \cite{zhang2021integrative},
the strain energy function is given as 
\begin{equation}\label{eq:new-muscle-energy}
	\begin{split}
	\mathfrak{W}  & =  \frac{a}{2b}\exp\left[b (I_1 - 3 )\right] - a \ln J  
	+ \frac{\lambda}{2}(\ln J)^{2} \\
	& + \sum_{i = f,s} \frac{a_i}{2b_i}\{\text{exp}\left[b_i\left(\mathit{I}_{ii}- 1 \right)^2\right] - 1\} \\
	& + \frac{a_{fs}}{2b_{fs}}\{\text{exp}\left[b_{fs}\mathit{I}^2_{fs} \right] - 1\} ,
	\end{split}
\end{equation}
where $a$, $b$, $a_f$, $b_f$, $a_s$, $b_s$, $a_{fs}$ and $b_{fs}$ 
represent eight positive material constants, 
and $\lambda$ is a Lamé parameter. 
The series of parameters $a$ have units of stress, 
while $b$ are dimensionless.
Here, 
the principle invariants are defined as
\begin{equation}\label{principle-invariants}
	I_1 = \operatorname{tr} \mathbb C, \quad I_2 = \frac{1}{2}\left[I^2_1 - \operatorname{tr}(\mathbb C^2)\right], \quad I_3 = \det(\mathbb C) = J^2,
\end{equation}
where the left Cauchy-Green deformation tensor 
$\mathbb C = \mathbb F^T \mathbb F$, 
and three other independent invariants 
associated with directional preferences are given by
\begin{equation}\label{extra-principle-invariants}
	I_{ff}  =   \mathbb C : \bm f^0 \otimes\bm f^0,
	\quad I_{ss}  =   \mathbb C : \bm s^0 \otimes\bm s^0,
	\quad I_{fs}  =   \mathbb C : \bm f^0 \otimes\bm s^0, 
\end{equation}
where $\bm f^0$ and $\bm s^0$  are the undeformed muscle fiber and sheet unit direction, respectively. 

The second Piola-Kirchhoff stress $\mathbb{S}$ can be derived by
\begin{equation}\label{eq:second-PK}
	\mathbb{S} = 2 \frac{\partial \mathfrak{W}}{\partial \mathbb{C}} -p\mathbb{C}^{-1} = 2\sum_{j} \frac{\partial \mathfrak{W}}{\partial \mathit{I}_j} \frac{\partial \mathit{I}_j}{\partial \mathbb{C}} -p\mathbb{C}^{-1} \quad j = 1, ff, ss, fs, 
\end{equation}
where 
\begin{equation}\label{eq:second-PK-2}
	\frac{\partial \mathit{I}_1}{\partial \mathbb{C}} =  \mathbb{I}, \quad
	 \frac{\partial \mathit{I}_{ff}}{\partial \mathbb{C}} = \bm{f}_0 \otimes \bm{f}_0, \quad
	 \frac{\partial \mathit{I}_{ss}}{\partial \mathbb{C}} = \bm{f}_0 \otimes \bm{f}_0, \quad
	 \frac{\partial \mathit{I}_{fs}}{\partial \mathbb{C}} = \bm{f}_0 \otimes \bm{s}_0 + \bm{s}_0 \otimes \bm{f}_0 ,
\end{equation}
and $p = \frac{\partial \mathfrak{W}}{\partial J}$ serves as the Lagrange multiplier introduced to enforce incompressibility. 
Substituting Eqs. \eqref{eq:new-muscle-energy} and \eqref{eq:second-PK-2} 
into Eq.\eqref{eq:second-PK} 
and applying $\fancy{$\tau$} = \mathbb{F}\mathbb{S}\mathbb{F}^{\operatorname{T}}$,
the Kirchhoff stress $\fancy{$\tau$}$ is obtained as
\begin{equation}
\begin{split}
	\fancy{$\tau$} & = \left\{ \lambda\ln J - a \right\} \mathbb{I} 
	+ a \, \text{exp} \left[b\left({\mathit{I}}_1 - 3 \right)\right] \fancy{$b$}\\ 
	& + 2a_f \left({\mathit{I}}_{f}- 1 \right)  \text{exp}\left[b_f\left({\mathit{I}}_{f}- 1 \right)^2\right] 
	\mathbb{F} (\bm{f}_0 \otimes \bm{f}_0) \mathbb{F}^{\operatorname{T}}  \\
	& + 2a_s \left({\mathit{I}}_{s}- 1 \right)  \text{exp}\left[b_s\left({\mathit{I}}_{s}- 1 \right)^2\right] 
	\mathbb{F} (\bm{s}_0 \otimes \bm{s}_0)  \mathbb{F}^{\operatorname{T}} \\
	& + a_fs {\mathit{I}}_{fs} \text{exp}\left[b_fs\left({\mathit{I}}_{fs}\right)^2\right] \mathbb{F}
	\left(\bm{f}_0 \otimes \bm{s}_0 + \bm{s}_0 \otimes \bm{f}_0\right)
	\mathbb{F}^{\operatorname{T}}.
\end{split}
\end{equation}

\subsection{Electrophysiologically induced active stress model}
Building on the methodology 
outlined in Refs \cite{nash2004electromechanical, zhang2021integrative}, 
we incorporate the stress tensor with the transmembrane potential $V_m$ 
using the active stress approach. 
This approach decomposes the Kirchhoff stress $\fancy{$\tau$}$ 
into passive and active components as
\begin{equation}
	\fancy{$\tau$} = \fancy{$\tau$}_{passive} + \fancy{$\tau$}_{active}, 
\end{equation}
where the passive component $\fancy{$\tau$}_{passive}$ 
describes the stress 
required to achieve a given passive muscle deformation, 
which is modeled by the above-mentioned Holzapfel-Odgen material, 
and the active component  $\fancy{$\tau$}_{active}$ 
denotes the tension activated by 
the depolarization of the propagating transmembrane potential. 
Following the active stress approach proposed in Ref.  \cite{nash2004electromechanical}, 
the active component is obtained as
\begin{equation}
	\fancy{$\tau$}_{active} 
	= T_a \mathbb{F} \bm{f}_0 \otimes \bm{f}_0 \mathbb{F}^{\operatorname{T}},
\end{equation}
where $T_a$ represents the active muscle contraction stress. 

\section{Methodology}\label{sec:SPHalgorithm}
\subsection{Total Lagrangian SPH}
Following Refs. \cite{monaghan2005smoothed, zhang2022review}, 
the momentum conservation Eq. \eqref{conservation_equation} is discretized
in the weak-form SPH approximation of the spatial derivative as
\begin{equation}\label{discrete_dynamic_equation}
	\rho_{i}^0 \bm{\ddot u}_i  = \sum\limits_j \left(\mathbb{P}_i {\mathbb{B}_i^0}^{\text{T}} + \mathbb{P}_j {\mathbb{B}_j^0}^{\text{T}} \right) \nabla_i ^0 W_{ij} V_j^0,
\end{equation}
where $\nabla_i^0 W_{ij}  = \frac{\partial W\left(\bm{r}_{ij}^0, h \right)}{\partial \bm{r}_{ij}^0} \bm{e}_{ij}^0 $
denotes the gradient of the kernel function evaluated 
at the initial reference configuration 
with $r_{ij}^0$ representing the initial particle distance 
and $\bm{e}_{ij}^0$ the initial unit vector pointing from particle $j$ to particle $i$.
Additionally, 
$\rho_{i}^0$ is the initial density of particle $i$, 
and $V_j^0$ is the initial volume of particle $j$. 
Here, the superscript $\left( \bullet \right)^0$ is introduced 
to represent variables defined at the initial reference configuration. 
The correction matrix $\mathbb{B}^0$ is adopted 
to ensure first-order completeness as \cite{randles1996smoothed, bonet2002simplified, vignjevic2009review}
\begin{equation} \label{eq:b-corection}
	\mathbb{B}_i^0  =  \left( 
	\sum\limits_j {V_j^0 \left( {\bm{r}_j^0  - \bm{r}_i^0 } \right) \otimes \nabla_i^0 W_{ij} }
	\right)^{-1}.
\end{equation}

The deformation tensor $\mathbb{F}$ is updated based on its rate of change,
which is approximated in the strong-form discretization   
of the spatial derivative \cite{monaghan2005smoothed, zhang2022review}  
as
\begin{equation}
	\label{deformation_tensor_change_rate}
	\frac{d\mathbb{F}_i}{dt}  = \dot{\mathbb{F}}_i = \sum\limits_j V_j^0 \left(\bm{\dot u}_j  - \bm{\dot u}_i \right) \otimes \nabla _i^0 W_{ij} \mathbb{B}_i^0.
\end{equation}
Following the approach in Ref. \cite{zhang2022artificial}, 
we introduce an artificial damping stress $\fancy{$\tau$}_d$ 
based on the Kelvin-Voigt type damper 
when calculating the Kirchhoff stress $\fancy{$\tau$}$ as
\begin{equation}
	\label{Kirchhoff_stress2}
	\fancy{$\tau$}_d = \frac{\chi}{2}  \frac{d \fancy{$b$}}{dt},
\end{equation}
where the artificial viscosity factor $\chi = \rho  c h/2$ with $c  = \sqrt {K/\rho  }$ and $h$ denoting the smoothing length, 
and the change rate of the left Cauchy-Green deformation gradient tensor
\begin{equation}
	\frac{d \fancy{$b$}}{dt} = \left[\frac{d\mathbb{F}}{dt}  \mathbb{F}^{\operatorname{T}} + \mathbb{F}  \left( \frac{d\mathbb{F}}{dt} \right)^{\operatorname{T}} \right].
\end{equation}

\subsection{Generalized essentially non-hourglass formulation}
\label{sec:hourglass}
Since the hourglass modes exhibit very large local, 
especially shear, deformation \cite{wu2023essentially}, 
we introduce a correction term to the discretization of 
shear-stress term to suppress this instability.
We first decompose the Kirchhoff stress by considering the material model 
aforementioned in Section \ref{sec:ConstitutiveRelation} as 
\begin{equation}
	\label{Kirchhoff_stress4}
	\fancy{$\tau$}= \fancy{$\tau$}_s
	+ \fancy{$\tau$}_r. 
\end{equation}
Here, the first term of the right-hand side 
$\fancy{$\tau$}_s = c\,\fancy{$b$}_e$, 
with $c = \left|\fancy{$b$}_e\right|^ {-\frac{1}{d}} G$ 
for elastic and plastic materials
and $c = a \, \text{exp} \left[b\left({\mathit{I}}_1 - 3 \right)\right]$
 for muscle model,
contains the main shear stress components, 
and the second term gives the remaining Kirchhoff stress. 
For example, 
$\fancy{$\tau$}_r= \frac{K}{2}\left( J^2 - 1 \right) \mathbb{I}  
	- \frac{1}{d} \left|\fancy{$b$}_e\right|^ {-\frac{1}{d}} G \operatorname{tr} \left( \fancy{$b$}_e \right) \mathbb{I} + \fancy{$\tau$}_d$ 
for the elastic and plastic materials applied in this study. 
Note that $\fancy{$b$}_e = \fancy{$b$}$ for elastic deformation, 
including those in the muscle models.

For standard elastic material, 
the shear part of the first Piola-Kirchhoff stress $\mathbb{P}_s = \fancy{$\tau$}_s \mathbb{F}^{-\operatorname{T}} 
= c \, \fancy{$b$} \mathbb{F}^{-\operatorname{T}}$, 
and the particle acceleration induced by the shear stress
can be obtained by the standard SPH method as follows
\begin{equation}\label{eq:SPH_method}
	\rho^0 \ddot {\bm{u}}_{s, i} = 
	\sum \limits_j 
\left(c_i \, \fancy{$b$}_i \mathbb{F}^{-T}_i
+ c_j \, \fancy{$b$}_j \mathbb{F}^{-T}_j\right) 
	\frac{\partial W\left(\bm{r}_{ij}^0, h \right)}{\partial \bm{r}_{ij}^0}V_j^0 
	\bm{e}_{ij}^0,
\end{equation}
which may suffer from serious hourglass modes. 
To obtain an essentially non-hourglass formulation 
as proposed in our previous study \cite{wu2023essentially}, 
the discretization for shear acceleration is instead obtained by 
applying a non-nested Laplacian formulation as
\begin{equation}\label{eq:Laplacian_formualtion}
	\rho_i^0 \ddot {\bm{u}}_{s, i} 
	= \sum \limits_j  \left(c_i + c_j\right) \frac {\bm{r}_{ij}}{r_{ij}^0}
	\frac{\partial W\left( r_{ij}^0,h \right)}{\partial r_{ij}^0 }
	V_j^0. 
\end{equation}
By using the entity $\mathbb{F}\mathbb{F}^{-1} = \mathbb{I}$ 
and $\fancy{$b$} = \mathbb{F}\mathbb{F}^{\operatorname{T}}$,
one can reformulate Eq. \eqref{eq:Laplacian_formualtion} 
approximately as
\begin{equation}\label{eq:Laplacian_formualtion-b}
\rho_i^0 \ddot {\bm{u}}_{s, i} 
= \sum \limits_j  \left(c_i \, \fancy{$b$}_i \mathbb{F}^{-T}_i
+ c_j \, \fancy{$b$}_j \mathbb{F}^{-T}_j\right)
\frac{\partial W\left( r_{ij}^0,h \right)}{\partial r_{ij}^0 }
V_j^0 \left[ \frac{1}{2}\left(\mathbb{F}_i^{-1} + \mathbb{F}_j^{-1}\right)
\frac{\bm{r}_{ij}}{r^0_{ij}} \right]. 
\end{equation}
Comparing Eqs. \eqref{eq:SPH_method} and \eqref{eq:Laplacian_formualtion-b}, 
one can observe that a tracing-back prediction of 
the initial inter-particle direction
from the current deformation gradient is
\begin{equation}
	\bm{e}_{ij}^0 \approx \frac{1}{2}\left(\mathbb{F}_i^{-1} + \mathbb{F}_j^{-1}\right)
	\frac{\bm{r}_{ij}}{r^0_{ij}}. 
\end{equation}
Such prediction is exact when the deformation is linear, 
but produces discrepancy for general, especially large deformations.  
Since Eq. \eqref{eq:SPH_method} is prone to hourglass modes and 
Eq. \eqref{eq:Laplacian_formualtion} essentially free of them, 
one can incorporate a correction term into Eq. \eqref{eq:SPH_method} 
based on the discrepancy
as 
\begin{equation}\label{eq:hourglass_control_elasticity}
	\rho^0 \ddot {\bm{u}}_{s, i} = 
	\sum \limits_j 
	\left(c_i \, \fancy{$b$}_i \mathbb{F}^{-T}_i
	+ c_j \, \fancy{$b$}_j \mathbb{F}^{-T}_j\right) 
	\frac{\partial W\left(\bm{r}_{ij}^0, h \right)}{\partial \bm{r}_{ij}^0}V_j^0 
	\left(\bm{e}_{ij}^0 
	+ \varphi \hat{\bm{e}}^0_{ij}
	\right),
\end{equation}
where
\begin{equation}\label{eq:discrepancy} 
\hat{\bm{e}}^0_{ij} = \frac{1}{2}\left(\mathbb{F}_i^{-1} + \mathbb{F}_j^{-1}\right)
\frac{\bm{r}_{ij}}{r^0_{ij}}
- \bm{e}^0_{ij}
\end{equation}
and $\varphi$ is a dimensionless modeling parameter 
adjusting the contribution of 
the reformulated essentially non-hourglass discretization of Eq. \eqref{eq:Laplacian_formualtion-b}.
If $\varphi$ degenerates to unity, 
Eq. \eqref{eq:hourglass_control_elasticity} 
recovers this reformulated non-hourglass discretization.

Note that the correction term is purely numerical,
and vanishes as 
the discrepancy decreases with 
increasing the resolution of discretization.
Also note that the correction term, 
being dimensionless and purely geometric, 
is independent of the material model 
which can be implemented in a straightforward way by simply extending 
$\fancy{$b$}$ to $\fancy{$b$}_e$ in Eq. \eqref{eq:hourglass_control_elasticity}. 
For example, $\fancy{$b$}_e$ can be obtained through return mapping as shown in Algorithm \ref{al:algorithm1} for plastic materials. 
Therefore, together with the
correction matrix $\mathbb{B}^0$ of Eq. \eqref{eq:b-corection} fulfilling the first-order completeness, 
the generalized essentially non-hourglass formulation can be written as
\begin{equation}\label{eq:hourglass_control_with_B}
\rho^0 \ddot {\bm{u}}_{s, i} = 
\sum \limits_j 
\left(
c_i \, \fancy{$b$}_{e,i} \mathbb{F}_i^{-\operatorname{T}} \mathbb{B}^0_i
+ c_j \, \fancy{$b$}_{e,j}\mathbb{F}_j^{-\operatorname{T}}
\mathbb{B}^0_j
\right) 
\frac{\partial W\left(\bm{r}_{ij}^0, h \right)}{\partial \bm{r}_{ij}^0}V_j^0 
\left(\bm{e}_{ij}^0 
+ \varphi \hat{\bm{e}}^0_{ij}
\right), 
\end{equation}
where the modeling parameter is proposed as
$\varphi = \alpha d \beta_{ij} \gamma_{ij}$. 
Here, $\beta_{ij} = W^0_{ij} / W_{0}$
leads to less correction to further neighbor particles, 
parameter $\alpha = 8.0$ according to the numerical experiment 
and remains constant throughout this work, 
and 
\begin{equation}
	\gamma_{ij} = \min \left(
	\max \left(\left|\hat{\bm{e}}^0_{ij}\right| - 0.05, 0\right), 
	1
	\right)
\end{equation}
serves as a magnitude limiter, 
allowing no correction when the discrepancy is less than 0.05 
and gradually increasing the correction 
until the discrepancy reaches a large magnitude of 1.05.  

\begin{figure}[htb!]
	\centering
	\includegraphics[trim = 2mm 6mm 2mm 2mm, width=0.85 \textwidth]{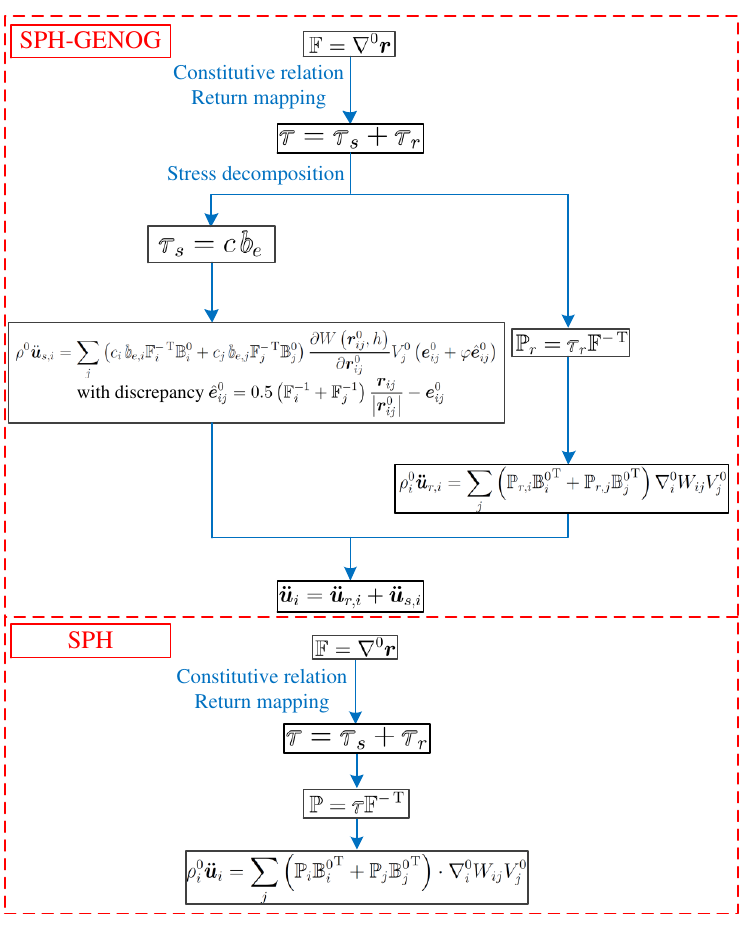}
	\caption{Flowcharts of the original total Lagrangian SPH (denoted as SPH) in Ref. \cite{han2018sph} 
		and present (denoted as SPH-GENOG) formulations.}
	\label{figs:solution_strategy}
\end{figure}

With $\mathbb{P}_r = \fancy{$\tau$}_r \mathbb{F}^{-\operatorname{T}}$ 
at hand,
the acceleration $\bm{\ddot u}_{r,i}$ of particle $i$, 
resulting from the remaining stress $\fancy{$\tau$}_{r,i}$, 
is calculated using the Eq. \eqref{discrete_dynamic_equation}
with $\mathbb{P}_r$ substituted for $\mathbb{P}$. 
Finally, the acceleration of particle $i$ is expressed as
\begin{equation}\label{acceleration}
	\bm{\ddot u}_i = \bm{\ddot u}_{r,i} + \bm{\ddot u}_{s,i}.
\end{equation}
For clarity, 
the flowcharts for the original and present SPH formulations are given, respectively, in Fig. \ref{figs:solution_strategy}. 

\subsection{Time integration scheme}\label{sec:time_integration}
In accordance with Ref. \cite{zhang2021multi}, 
the position-based Verlet scheme is employed for time integration. 
Initially, 
the deformation gradient tensor, density, and particle position 
are updated to the midpoint of $n$-th time step as
\begin{equation}\label{eq:verlet-first-half-solid}
\begin{cases}
\mathbb{F}^{n + \frac{1}{2}} = \mathbb{F}^{n} + \frac{1}{2} \Delta t \dot {\mathbb{F}}^n\\
\rho^{n + \frac{1}{2}} = \rho^0 \frac{1}{J} \\
\bm{r}^{n + \frac{1}{2}} = \bm{r}^n + \frac{1}{2} \Delta t \bm{\dot u}^n.
\end{cases}
\end{equation}
Upon calculating the Kirchhoff stress $\fancy{$\tau$}$ 
based on the applied constitutive relation 
and subsequently obtaining particle acceleration using Eq. \eqref{acceleration}, 
the velocity is updated through
\begin{equation}\label{eq:verlet-first-mediate-solid}
\bm{\dot u}^{n + 1} = \bm{\dot u}^{n} +  \Delta t  \bm{\ddot u}^{n+1}.
\end{equation}
After that, 
the rate of change of the deformation gradient tensor 
$\dot {\mathbb{F}}^{n+1}$ is computed 
using Eq. \eqref{deformation_tensor_change_rate}. 
Finally, the deformation gradient tensor and particle positions 
are updated to a new time step with
\begin{equation}\label{eq:verlet-second-final-solid}
\begin{cases}
\mathbb{F}^{n + 1} = \mathbb{F}^{n + \frac{1}{2}} + \frac{1}{2} \Delta t \dot {\mathbb{F}}^{n+1}\\
\rho^{n + 1} = \rho^0 \frac{1}{J} \\
\bm{r}^{n + 1} = \bm{r}^{n + \frac{1}{2}} + \frac{1}{2} \Delta t \bm{\dot u}^{n + 1}.
\end{cases}
\end{equation}
To maintain the numerical stability, the time step $\Delta t$ 
is given by the standard criteria
\begin{equation}\label{eq:dt}
	\Delta t   =  \text{CFL} \min\left(\frac{h}{c_v + |\bm{\dot u}|_{max}},
	\sqrt{\frac{h}{|\bm{\ddot u}|_{max}}} \right).
\end{equation} 
Note that the present Courant-Friedrichs-Lewy (CFL) number is set as $0.6$.

\section{Numerical examples}\label{sec:examples}
In this section, 
we conduct a series of benchmark tests with available analytical 
or numerical reference data from the literature 
to qualitatively and quantitatively assess the accuracy and stability of 
the proposed generalized essentially non-hourglass formulation 
(denoted as SPH-GENOG). 
For comparison, we also consider the original standard SPH formulation. 
Following the validation, 
we explore the deformation of a complex problem of Oobleck octopus 
to showcase the potential of the present formulation. 
The $5th$-order Wendland kernel \cite{wendland1995piecewise}, 
characterized by a smoothing length of $h = 1.15dp$ (where $dp$ denotes the initial particle spacing) and a cut-off radius of $2.3dp$, 
is employed throughout.

\subsection{Oscillating plate}
\begin{figure}[htb!]
	\centering
	\includegraphics[trim = 0mm 6mm 2mm 4mm, width=0.6 \textwidth]{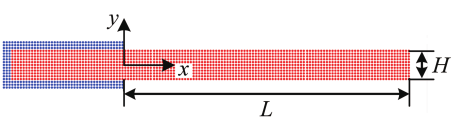}
	\caption{Oscillating plate: Initial configuration.}
	\label{figs:oscillating_plate_setup}
\end{figure}

First, 
we examine the oscillation of a thin plate with one edge fixed 
while the other edges remain free. 
This classical problem has been extensively explored 
both theoretically \cite{landau1986course} 
and numerically \cite{zhang2017generalized, wu2023essentially} 
in the literature.
The problem is represented as a plane strain scenario, 
modeling a 2D plate strip of length $L = 0.2~\text{m}$, 
perpendicular to the fixed edge, 
with a thickness of $H = 0.02~\text{m}$.
In accordance with previous studies \cite{zhang2017generalized, wu2023essentially}, 
the plate strip is clamped between several layers of constrained SPH particles, 
as depicted in Figure \ref{figs:oscillating_plate_setup}.
The initial velocity, 
denoted as $v_y$ and directed perpendicular to the plate strip, 
is prescribed as follows
\begin{equation}
v_y(x) = v_f c \frac{f(x)}{f(L)},
\end{equation}
where $v_f$ represents a constant that varies among different cases, 
and 
\begin{equation}
	\begin{split}
	f(x) &= \left(\sin(kL) + \sinh(kL) \right) \left(\cos(kx) - \cosh(kx) \right) \\
              & - \left(\cos(kL) + \cosh(kL) \right) \left(\sin(kx) - \sinh(kx) \right)
	\end{split}
\end{equation}
with $k$ determined by 
\begin{equation}
	\cos(kL) \cosh(kL) = -1
\end{equation}
and $kL = 1.875$. 
The material properties are defined as follows: 
density $\rho_0 = 1000.0 ~\text{kg} / \text{m}^3$, 
Young's modulus $E = 2.0~\text{MPa}$ and Poisson's ratio $\nu$ varies for different cases. 
The theoretical expression for the frequency $\omega$ 
of the oscillating plate is provided by
\begin{equation}
	\omega ^2 = \frac{E H^2 k^4}{12 \rho \left(1 - \nu^2 \right)}.
\end{equation}
\begin{figure}[htb!]
	\centering
	\includegraphics[trim = 2mm 9mm 2mm 2mm, width=\textwidth] {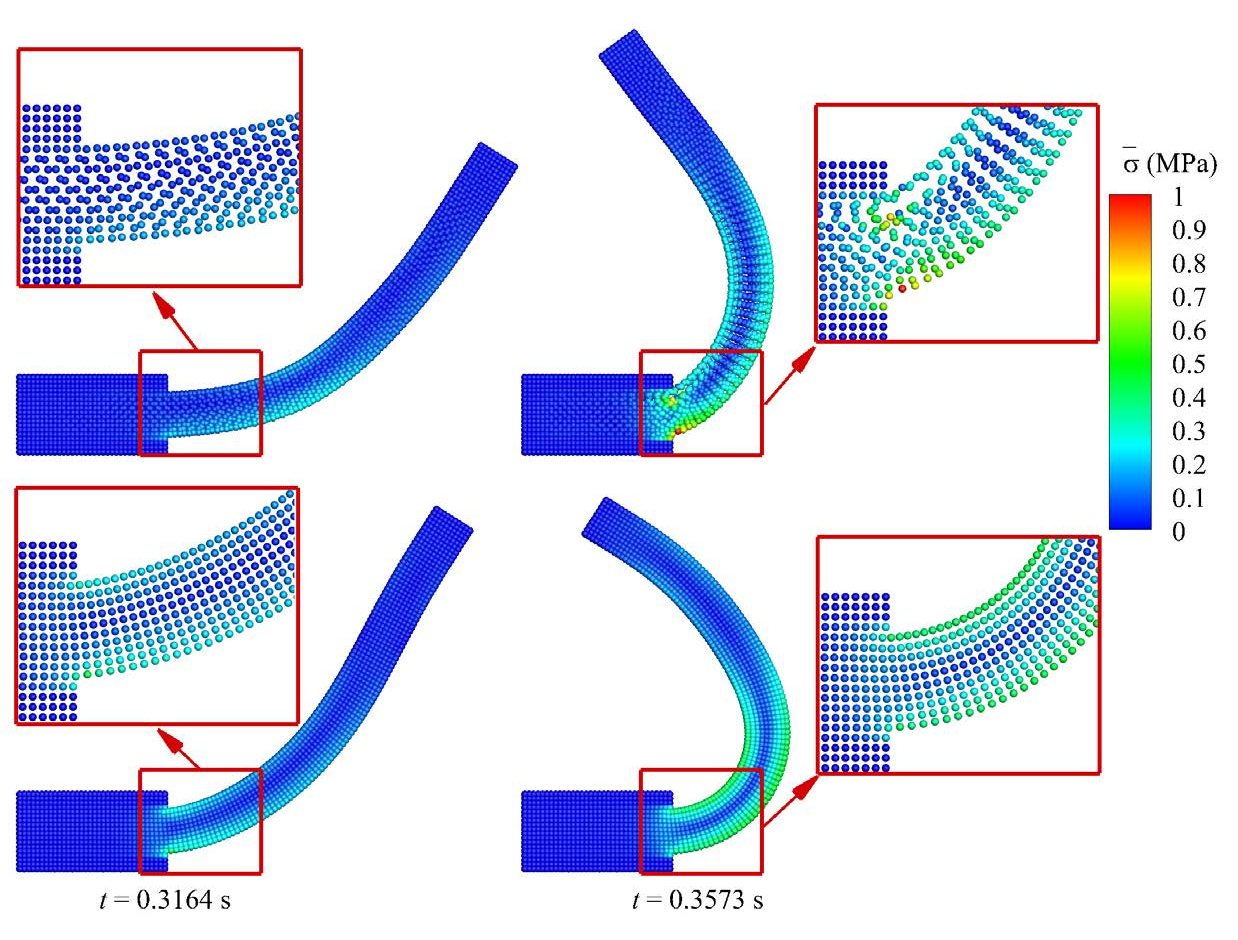}
	\caption{Oscillating plate: Comparison of the deformed configuration colored by von Mises stress $\bar\sigma$ at two time instants obtained by the SPH (top panel) and SPH-GENOG (bottom panel) 
		with the length $L = 0.2~\text{m}$, height $H = 0.02~\text{m}$, $v_f = 0.15$, and spatial particle discretization $H / dp = 10$. 
		The material is modeled with density $\rho_0 = 1000.0 ~\text{kg} / \text{m}^3$, Young's modulus $E = 2.0~\text{MPa}$, and Poisson's ratio $\nu = 0.3975$. }
	\label{figs:oscillating_plate_comparison}
\end{figure}

Figure \ref{figs:oscillating_plate_comparison} depicts the deformed particle configuration, 
accompanied by the von Mises stress $\bar\sigma$ contour, 
simulated by both SPH and SPH-GENOG 
under $v_f = 0.15$ 
and Poisson's ratio $\nu = 0.3975$.
It can be noted that,
the SPH results exhibit particle disorder under large deformation,
evident in top panel of Fig. \ref{figs:oscillating_plate_comparison}, 
especially in the vicinity of maximum stress.
As the plate strip undergoes larger deformation, 
an increasing number of particle pairs adhere together.
In contrast, 
SPH-GENOG mitigates such instability, 
exhibiting smooth deformation and stress fields.
\begin{figure}[htb!]
	\centering
	\includegraphics[trim = 2mm 6mm 2mm 2mm, width=\textwidth] {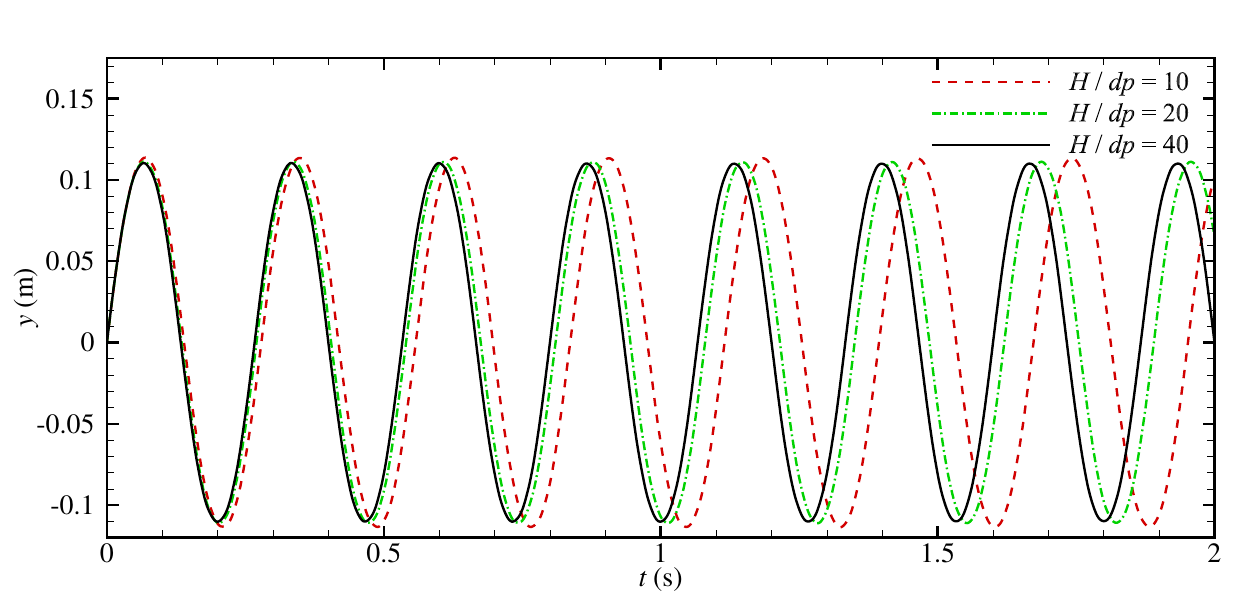}
	\caption{Oscillating plate: Time history of the vertical position $y$ 
		observed at the midpoint of the plate strip end obtained by the SPH-GENOG 
		in the long term
		with the length $L = 0.2~\text{m}$, height $H = 0.02~\text{m}$ and $v_f = 0.05$.
		The material is modeled with density $\rho_0 = 1000.0 ~\text{kg} / \text{m}^3$, Young's modulus $E = 2.0~\text{MPa}$ and Poisson's ratio $\nu = 0.3975$. 
		Note that $dp$ is the initial particle spacing.}
	\label{figs:oscillating_plate_convergence}
\end{figure}
\begin{table}[htb!]
	\centering
	\caption{Oscillating plate: Quantitative validation of the oscillation period for $L = 0.2~\text{m}$ and $H = 0.02~\text{m}$ with various $v_f $ and $\nu$.}
	\begin{tabular}{ccccc}
		\hline
		$v_f $   & $\nu$  & $T_\text{SPH-GENOG}$ & $T_\text{Theoretical}$ & Error \\ 
		\hline
		0.05 			& 0.22  	& 0.29355     & 0.27009      & 8.69\%\\
		0.1	 		     & 0.22      & 0.29261     & 0.27009      & 8.34\%\\
		0.15	 		& 0.22      & 0.29202     & 0.27009      & 8.12\%\\
		~\\
		0.05 			& 0.30  	& 0.28201     & 0.26412      & 6.77\%\\
		0.1	 		     & 0.30      & 0.28088      & 0.26412      & 6.35\%\\
		0.15	 		& 0.30      & 0.28079      & 0.26412      & 6.31\%\\
		~\\
		0.05 			& 0.40  	& 0.26634      & 0.25376      & 4.96\%\\
		0.1	 		     & 0.40      & 0.26515     & 0.25376      & 4.49\%\\
		0.15	 		& 0.40      & 0.26796     & 0.25376      & 5.60\%\\
		\hline	
	\end{tabular}
	\label{tab:oscillating_plate_period1}
\end{table}
\begin{table}[htb!]
	\centering
	\caption{Oscillating plate: Quantitative validation of the oscillation period for $L = 0.2~\text{m}$ and $H = 0.01~\text{m}$ with various $v_f $ and $\nu$.}
	\begin{tabular}{ccccc}
		\hline
		$v_f $   & $\nu$  & $T_\text{SPH-GENOG}$ & $T_\text{Theoretical}$ & Error \\ 
		\hline
		0.05 			& 0.22  	& 0.56932    & 0.54018      & 5.39\%\\
		0.1	 		     & 0.22      & 0.56229     & 0.54018      & 4.09\%\\
		0.15	 		& 0.22      & 0.56048     & 0.54018      & 3.76\%\\
		~\\
		0.05 			& 0.30  	& 0.54557    & 0.52824      & 3.28\%\\
		0.1	 		     & 0.30      & 0.54006     & 0.52824      & 2.24\%\\
		0.15	 		& 0.30      & 0.53472     & 0.52824      & 1.23\%\\
		~\\
		0.05 			& 0.40  	& 0.51246     & 0.50752      & 0.97\%\\
		0.1	 		     & 0.40      & 0.50796     & 0.50752      &0.09\%\\
		0.15	 		& 0.40      & -     & -      & -\\
		\hline	
	\end{tabular}
	\label{tab:oscillating_plate_period2}
\end{table}

In order to validate the accuracy of present formulation, 
a convergence study 
and comparisons between numerical and theoretical solutions are undertaken. 
 The convergence study involves testing three distinct spatial resolutions: 
 $H / dp =10$, $H / dp =20$, and $H / dp =40$. 
The time history of vertical position $y$ of the midpoint at the end of the strip, 
with $v_f = 0.05$,
is illustrated in Figure \ref{figs:oscillating_plate_convergence}.
It can be observed that the discrepancies among various solutions 
diminish rapidly as the spatial resolution increases. 
Furthermore, a long-term simulation is conducted herein 
to underscore the numerical stability of the proposed formulation.

For quantitative validation, 
oscillation period $T$ obtained by the present SPH-GENOG 
with a spatial particle resolution of $H / dp = 40$ 
are presented in Table \ref{tab:oscillating_plate_period1}. 
A comparison is made with theoretical solutions 
across a broad range of $v_f$ and $\nu$.
The error remains below 9.00\% for $\nu = 0.22$ 
and decreases to approximately 5.00\% 
as the Poisson's ratio is increased to 0.4.
Considering the assumption of a very small thickness in the analytical theory,
Table \ref{tab:oscillating_plate_period2} presents a comparison 
where the length $L$ remains the same
while the thickness $H$ is reduced to half of its previous value.
A significantly improved agreement is achieved, 
with the maximum error decreasing to less than 1.0\% for $\nu = 0.4$.
Noted that when $v_f = 0.15$ and $\nu = 0.4$, 
the deformation becomes substantial, 
leading to the plate coming into contact with the constrained base. 
Consequently, the period of the plate in this scenario is not informative.

\subsection{Bending column}
\begin{figure}[htb!]
	\centering
	\includegraphics[trim = 0mm 6mm 2mm 4mm]{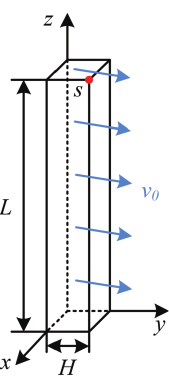}
	\caption{Bending column: Initial configuration.}
	\label{figs:bending_column_setup}
\end{figure}

To further assess the robustness and accuracy of the present formulation, 
we address a bending-dominated problem 
with a pre-existing numerical solution available 
in the literature \cite{aguirre2014vertex} for quantitative validation.
Both neo-Hookean and Holzapfel-Odgen material models 
are employed in this investigation.
Following Ref. \cite{zhang2021integrative}, 
a column with a length of $L = 6 \operatorname{m}$ and a square cross-section (height $H = 1 \operatorname{m}$) is clamped at its bottom, 
oscillating freely under the imposition of an initial uniform velocity
$\bm{v_0} = 10\left(\frac{\sqrt{3}}{2}, \frac{1}{2}, 0\right)^{\operatorname{T}}$, 
as depicted in Fig. \ref{figs:bending_column_setup}.
The neo-Hookean material is modeled 
with density $\rho_0 = 1100 ~\text{kg} / \text{m}^3$,  
Young's modulus $E = 17 ~\text{MPa}$, and Poisson's ratio $\nu = 0.45$. 
For the Holzapfel-Odgen model, 
material parameters are detailed in Table \ref{tab:Holzapfel_Ogden_material} with anisotropic terms adjusting accordingly. 
It is important to note that the Poisson's ratio $\nu$ of the Holzapfel-Odgen material 
is also set as 0.45 and $a = E / 2(1 + \nu)$ 
to facilitate a direct comparison with neo-Hookean in the isotropic scenario.
\begin{table}[htb!]
	\centering
	\caption{Bending column: Parameters for the Holzapfel-Ogden material model. Note that the anisotropic terms are set to zero for the isotropic material.}
	\begin{tabular}{cccc}
		\hline
		$a = 5.86$ MPa   & $a_f = ka$  & $a_s = 0.0 $ & $a_{fs} = 0.0$ \\ 
		\hline
		$b = 1.0$   & $b_f = 0.0$  & $b_s = 0.0 $ & $b_{fs} = 0.0$ \\ 
		\hline	
	\end{tabular}
	\label{tab:Holzapfel_Ogden_material}
\end{table}
\begin{figure}[htb!]
	\centering
	\includegraphics[trim = 2mm 8mm 2mm 2mm, width=\textwidth]{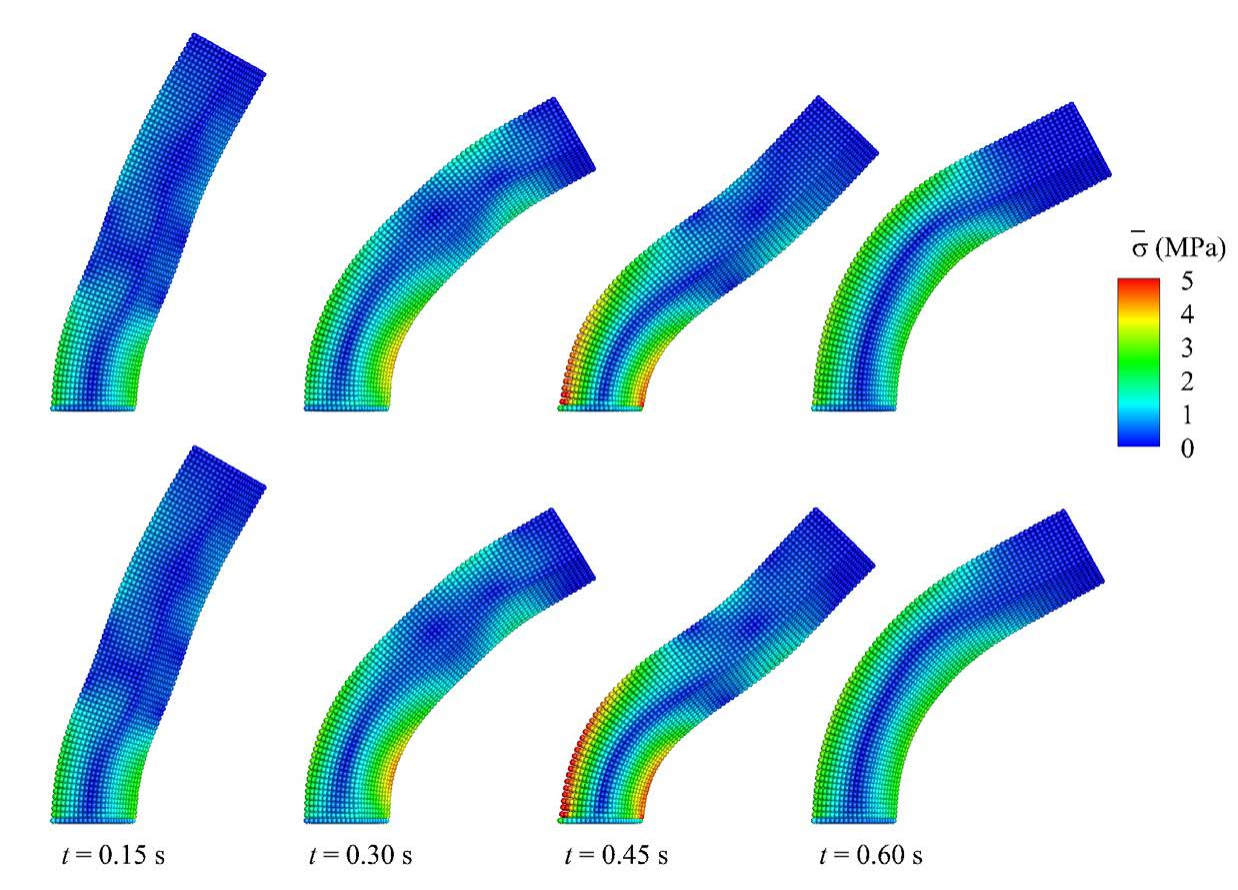}
	\caption{Bending column: Deformed configuration colored by von Mises stress $\bar\sigma$ at serial temporal instants 
		for Neo-Hookean (top panel) and Holzapfel-Ogden (bottom panel) materials 
		obtained by the present SPH-GENOG with initial uniform velocity $\bm{v_0} = 10\left(\frac{\sqrt{3}}{2}, \frac{1}{2}, 0\right)^{\operatorname{T}} ~\text{m/s}$. 
		The spatial particle discretization is set as $H / dp =12$ with $H$ denoting the height of the column and $dp$ the initial particle spacing.}
	\label{figs:bending_column_v10}
\end{figure}
\begin{figure}[htb!]
	\centering
	\includegraphics[trim = 0mm 6mm 2mm 2mm, width=\textwidth]{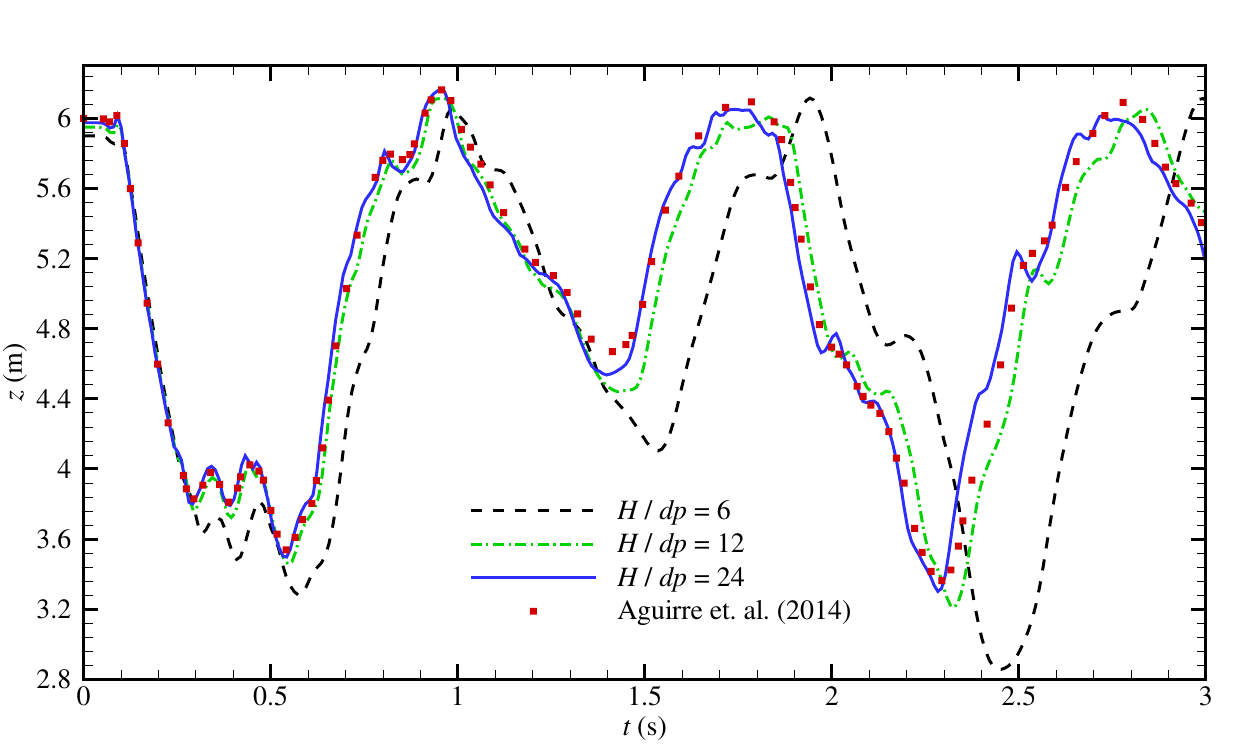}
	\caption{Bending column: Time history of the vertical position $z$ observed at node $S$ obtained by the SPH-GENOG for isotropic Holzapfel-Ogden material with three different spatial resolutions and the initial uniform velocity $\bm{v_0} = 10\left(\frac{\sqrt{3}}{2}, \frac{1}{2}, 0\right)^{\operatorname{T}} ~\text{m/s}$,
		and its comparison with that of Aguirre et al. \cite{aguirre2014vertex}. 
		}
	\label{figs:bending_column_convergence}
\end{figure}
\begin{figure}[htb!]
	\centering
	\includegraphics[trim = 2mm 10mm 2mm 2mm, width=\textwidth] {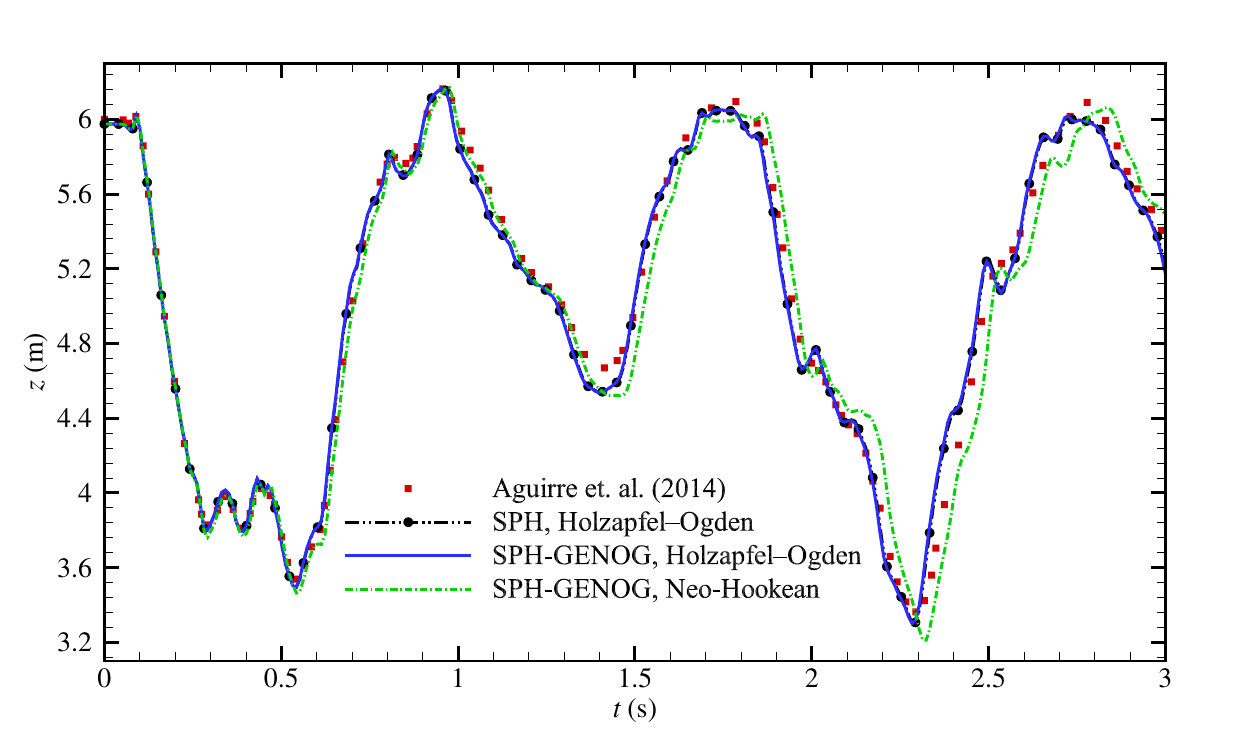}
	\caption{Bending column: Time history of the vertical position $z$ observed at node $S$ obtained by the SPH-GENOG and SPH with initial uniform velocity $\bm{v_0} = 10\left(\frac{\sqrt{3}}{2}, \frac{1}{2}, 0\right)^{\operatorname{T}} ~\text{m/s}$, 
		and its comparison with that of Aguirre et al. \cite{aguirre2014vertex}. 
		The spatial particle discretization is $H / dp = 24$.}
	\label{figs:bending_column_comparison}
\end{figure}
\begin{figure}[htb!]
	\centering
	\includegraphics[trim = 2mm 10mm 2mm 2mm, width=\textwidth] {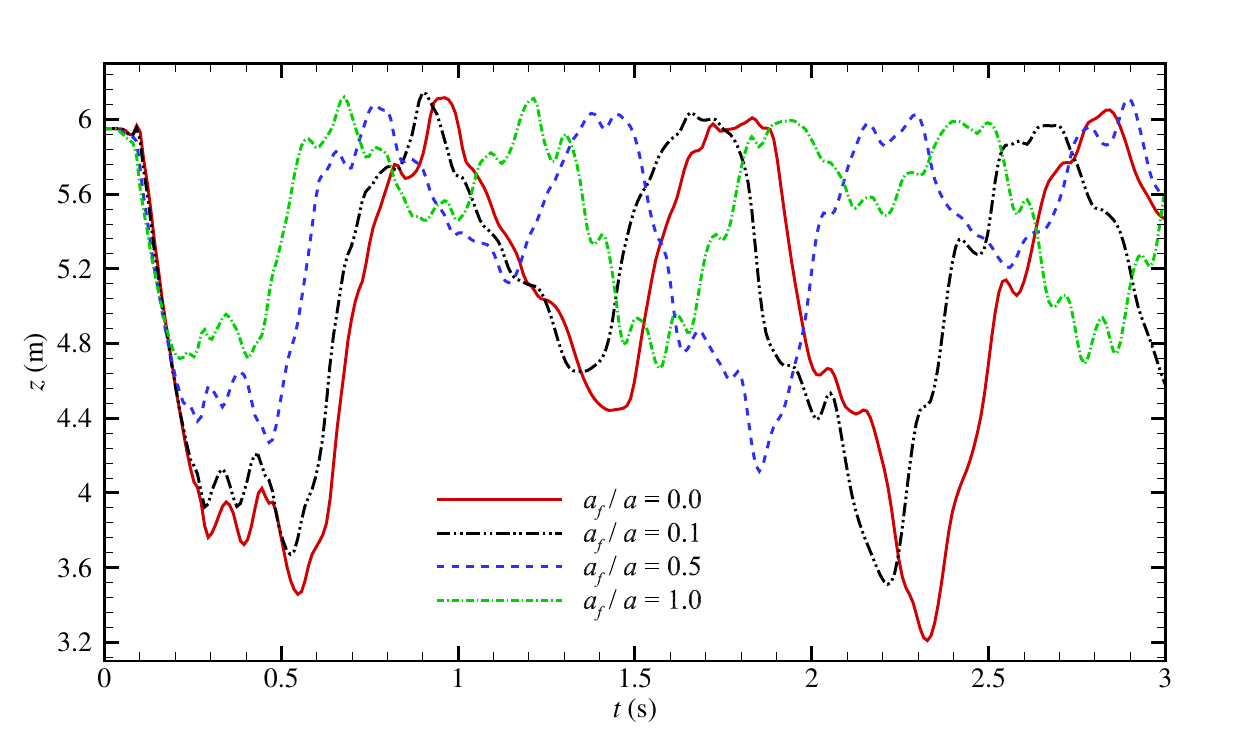}
	\caption{Bending column: Time history of the vertical position $z$ observed at node $S$ obtained by the SPH-GENOG
		for the Holzapfel-Ogden material model with initial uniform velocity $\bm{v_0} = 10\left(\frac{\sqrt{3}}{2}, \frac{1}{2}, 0\right)^{\operatorname{T}} ~\text{m/s}$.
		The spatial particle discretization is $H / dp = 12$.}
	\label{figs:bending_column_comparison_aniso}
\end{figure}
\begin{figure}[htb!]
	\centering
	\includegraphics[trim = 0mm 8mm 2mm 4mm, width=\textwidth]{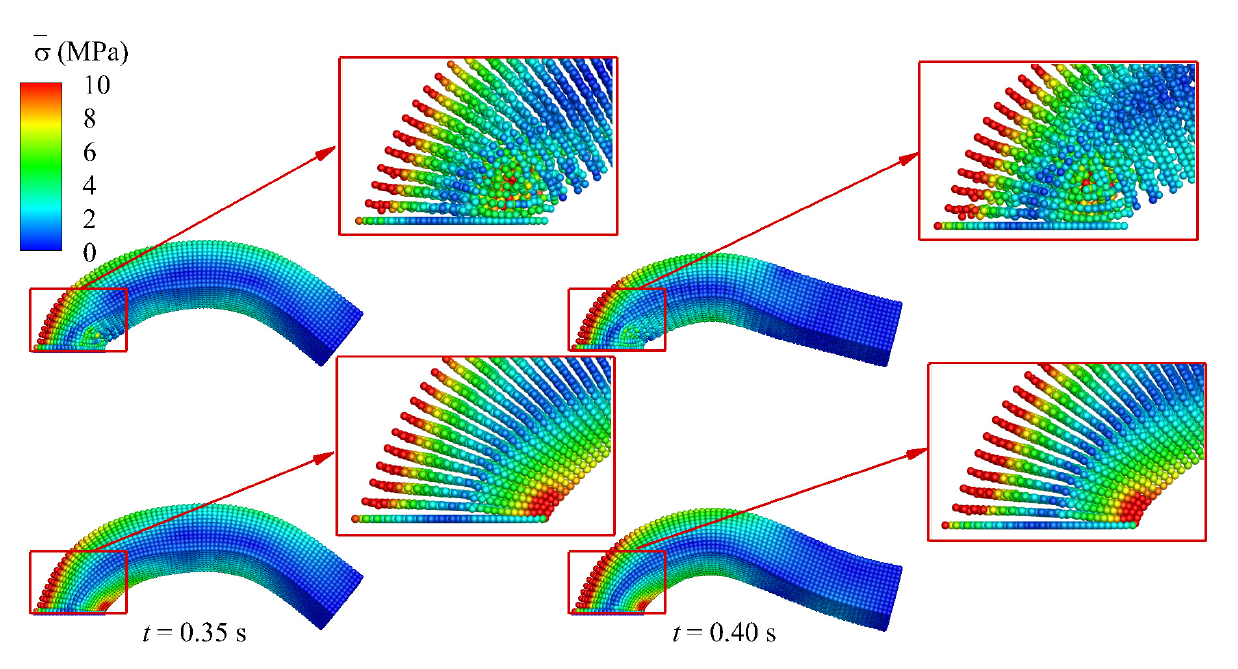}
	\caption{Bending column:  Deformed configuration colored by von Mises stress at two temporal instants obtained by the SPH (top panel) and SPH-GENOG (bottom panel) for isotropic Holzapfel-Ogden material model with initial uniform velocity $\bm{v_0} = 20\left(\frac{\sqrt{3}}{2}, \frac{1}{2}, 0\right)^{\operatorname{T}} ~\text{m/s}$. 
		The spatial particle discretization is $H / dp = 12$.}
	\label{figs:bending_column_v20}
\end{figure}
\begin{figure}[htb!]
	\centering
	\includegraphics[trim = 0mm 6mm 2mm 4mm, width=\textwidth]{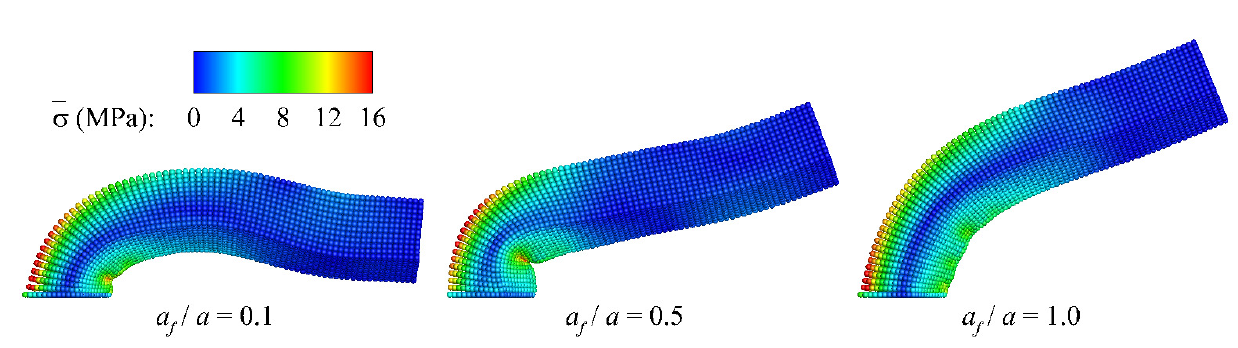}
	\caption{Bending column:  Deformed configuration at 0.4 s colored by von Mises stress obtained by the SPH-GENOG for anisotropic Holzapfel-Ogden material model with initial uniform velocity $\bm{v_0} = 20\left(\frac{\sqrt{3}}{2}, \frac{1}{2}, 0\right)^{\operatorname{T}} ~\text{m/s}$. 
		The spatial particle discretization is $H / dp = 12$.}
	\label{figs:bending_column_v20_stress}
\end{figure}

Figure \ref{figs:bending_column_v10} illustrates the time evolution 
of the deformed configuration,
represented by the von Mises stress contour,
as obtained through the present formulation.
The results obtained from both material models exhibit remarkable similarity, featuring a well-ordered particle distribution and a smooth stress field.
For quantitative validation, 
the time history of the $z$-axis position of point $S$ 
(as marked in Fig. \ref{figs:bending_column_setup}) 
for the isotropic Holzapfel-Ogden material model 
is presented in Fig. \ref{figs:bending_column_convergence}.
Different spatial resolutions, 
$H/dp = 6$, $H/dp = 12$, and $H/dp = 24$,
are considered, 
with a comparison to the reference results 
reported by Aguirre et al. \cite{aguirre2014vertex}.
Notably, 
robust convergence and a high level of agreement 
are evident with increasing spatial resolution.
As shown in Fig. \ref{figs:bending_column_comparison}, 
the outcomes computed by SPH and SPH-GENOG 
closely align with negligible discrepancies, 
and the quantitative disparities between the two materials are also minimal.

We further show the versatility of the present formulation 
by investigating this example
incorporating the anisotropic Holzapfel-Odgen material model. 
In the case of anisotropic material, 
we set the fiber and sheet directions aligned with $z$ and $x$ coordinates, respectively. 
We conduct three tests with varying anisotropic ratios: 
$a_f /a = 0.1$, $a_f /a = 0.5$, and $a_f /a = 1.0$. 
Figure \ref{figs:bending_column_comparison_aniso} shows 
the time history of the vertical displacement of point S. 
It can be observed that the deformation is reduced as the anisotropic ratio increases. 

To showcase the superior performance of the present formulation, 
we tackle a more challenging problem by elevating the initial velocity to 
$\bm{v_0} = 20\left(\frac{\sqrt{3}}{2}, \frac{1}{2}, 0\right)^{\operatorname{T}} ~\text{m/s}$
for the Holzapfel-Ogden material model.
As shown in Fig. \ref{figs:bending_column_v20}, 
the simulation result of the SPH exhibits noticeable particle disorder, 
especially near the clamped bottom where the maximum von Mises stress exists,
while the present SPH-GENOG captures 
a highly regular particle distribution and a smooth stress field, 
underscoring the robustness of the proposed formulation.
Furthermore, 
Fig. \ref{figs:bending_column_v20_stress} presents the deformed configuration 
colored by von Mises stress for the anisotropic Holzapfel-Odgen material model.

\subsection{Twisting column}
\begin{figure}[htb!]
	\centering
	\includegraphics[trim = 0mm 6mm 2mm 4mm]{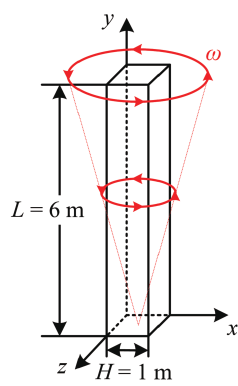}
	\caption{Twisting column: Initial configuration.}
	\label{figs:twsting_column_setup}
\end{figure}

In this section, 
we extend the bending column problem to encompass a twisting column, 
following Refs. \cite{lee2016new, lee2019total, zhang2022artificial}.
As illustrated in Fig. \ref{figs:twsting_column_setup}, 
the twisting is initiated with a sinusoidal rotational velocity field given by $\bm{\omega} = \left[ 0, \Omega_0 \operatorname{sin}\left( \pi y_0 /2 L \right),  0\right]$ with $\Omega_0 = 105~\operatorname{rad/s}$.
The column, 
modeled using the Holzapfel-Odgen material, 
is assumed to exhibit nearly incompressible behavior, 
with a Poisson's ratio of $\nu = 0.499$. 
The remaining material parameters remain consistent 
with those outlined in the previous section.
\begin{figure}[htb!]
	\centering
	\includegraphics[trim = 2mm 6mm 2mm 2mm, width=\textwidth]{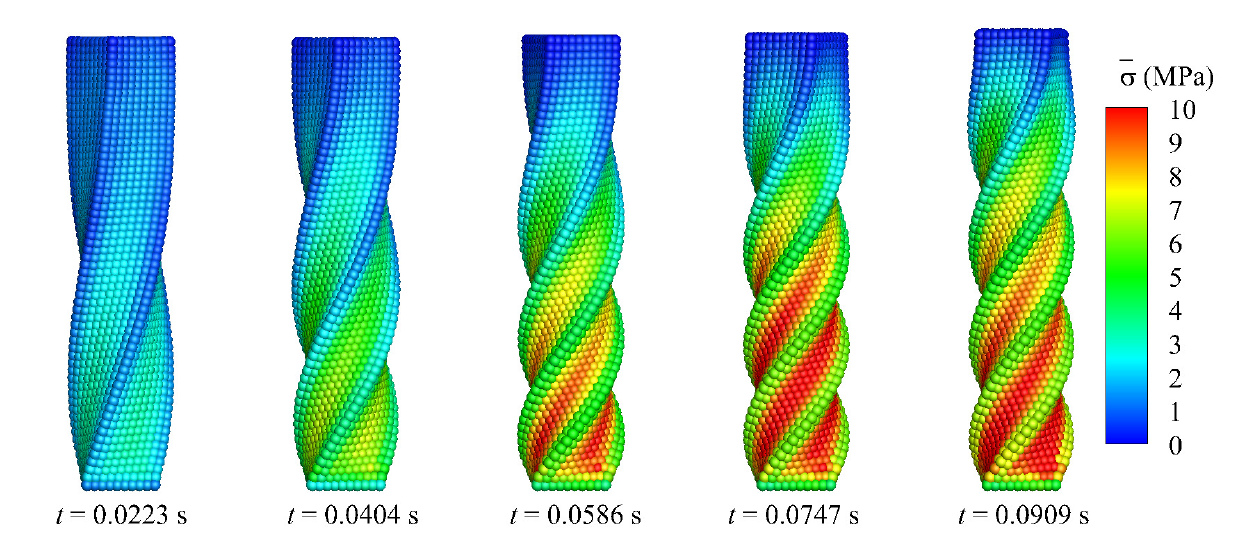}
	\caption{Twisting column: Deformed configuration colored by von Mises stress at different time instants for the isotropic Holzapfel-Odgen material model obtained by SPH-GENOG. 
		The initial rotational velocity $\bm{\omega} = \left[ 0, \Omega_0 \operatorname{sin}\left( \pi y_0 / 2 L \right),  0\right]$ with $\Omega_0 = 105~\operatorname{rad/s}$. 
		The spatial particle discretization is set as $H / dp = 10$ with $H$ denoting the height of the column and $dp$ the initial particle spacing.}
	\label{figs:twsting_column_w105}
\end{figure}
\begin{figure}[htb!]
	\centering
	\includegraphics[trim = 2mm 12mm 2mm 2mm, width=\textwidth]{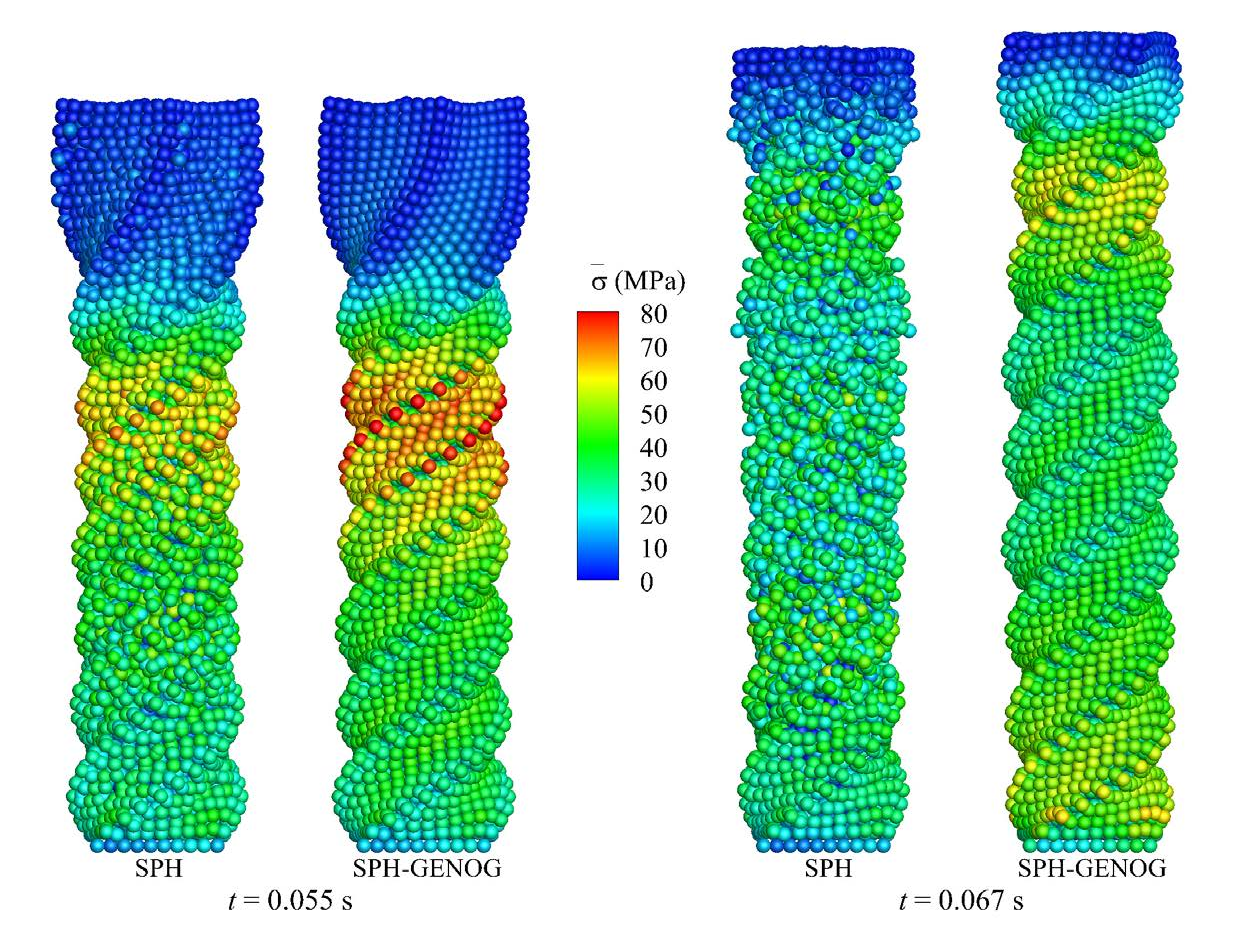}
	\caption{Twisting column: Deformed configuration colored by von Mises stress at two time instants for the isotropic Holzapfel-Odgen material model obtained by SPH and SPH-GENOG with initial sinusoidal rotational velocity $\Omega_0 = 330~\operatorname{rad/s}$. 
		The spatial particle discretization is set as $H / dp = 10$.}
	\label{figs:twsting_column_w330}
\end{figure}
\begin{figure}[htb!]
	\centering
	\includegraphics[trim = 4mm 6mm 4mm 4mm, width=\textwidth]{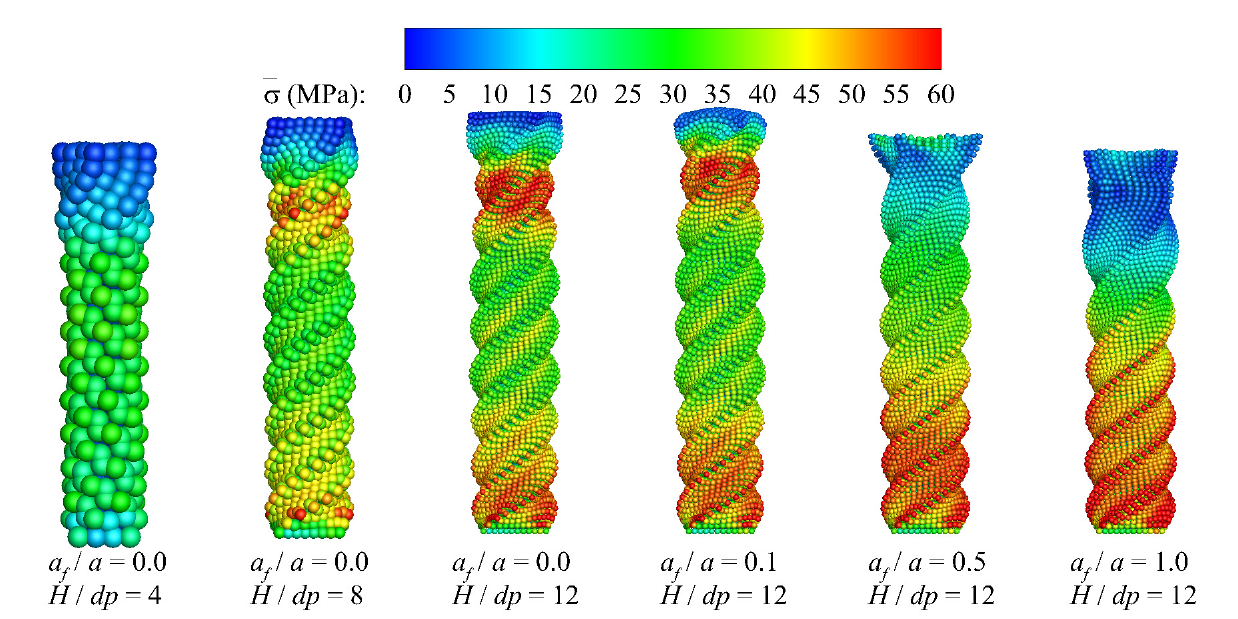}
	\caption{Twisting column: A sequence of particle refinement and anisotropic
		ratio increasing analyses using the SPH-GENOG with the initial sinusoidal rotational velocity $\Omega_0 = 330~\operatorname{rad/s}$.}
	\label{figs:twsting_column_w330_convergence}
\end{figure}
\begin{figure}[htb!]
	\centering
	\includegraphics[trim = 0mm 4mm 0mm 0mm, width=\textwidth]{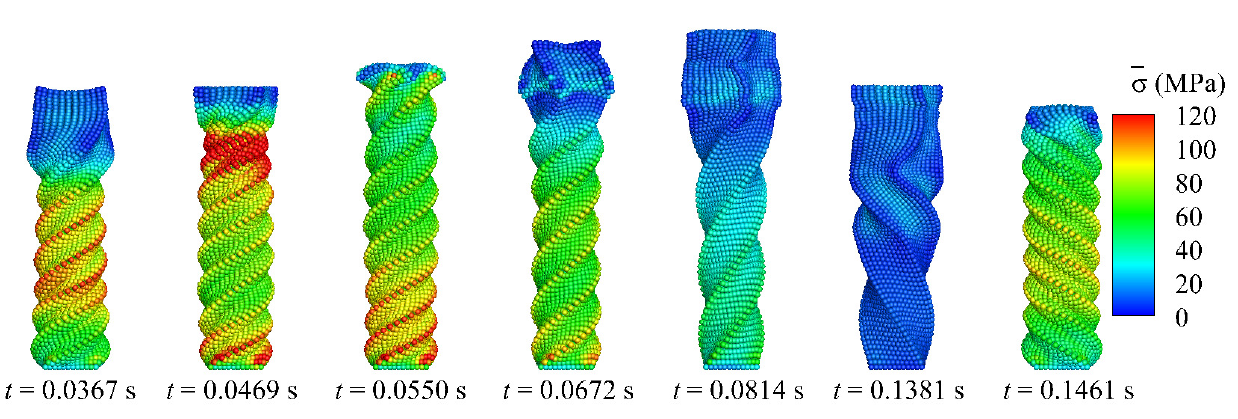}
	\caption{Twisting column: Deformed configuration plotted with von Mises stress at serial time instants when $a_f / a = 1.0$ obtained by the SPH-GENOG with the initial sinusoidal rotational velocity $\Omega_0 = 480~\operatorname{rad/s}$. 
		The spatial particle discretization is set as $H / dp = 12$.}
	\label{figs:twsting_column_w480}
\end{figure}
Figure \ref{figs:twsting_column_w105} presents 
the deformed configuration of the isotropic material model
at different time instants, 
accompanied by the von Mises stress contour obtained through SPH-GENOG.
The simulation performs well, 
exhibiting deformation patterns highly consistent with 
those reported in the literature (see Fig. 28 in Ref. \cite{lee2016new}). 
Addressing a notably more challenging scenario, 
we increase the initial angular velocity to
$\Omega_0 = 330~\operatorname{rad/s}$ with a Poisson’s ratio of $\nu = 0.49$.
As shown in Fig. \ref{figs:twsting_column_w330}, 
the proposed formulation demonstrates stability, 
contrasting with the visibly disordered particle results from SPH.
Conducting a convergence study involving sequential refinement of spatial resolution from $H/dp = 4$ to $H/dp = 8$ and $H/dp = 12$,
and an analysis of anisotropic behavior across varying anisotropic ratios 
($a_f /a = 0.1$, $a_f /a = 0.5$, and $a_f /a = 1.0$), 
Fig. \ref{figs:twsting_column_w330_convergence}
demonstrates robust convergence properties for both deformation and von Mises stress $\bar\sigma$, 
and provides insight into the smooth stress characteristic 
of the anisotropic Holzapfel-Odgen material.

Finally, to further assess the robustness of the present formulation,
we increase the initial angular velocity 
to $\Omega_0 = 480~\operatorname{rad/s}$
with $a_f / a = 1.0$.
As shown in Figure \ref{figs:twsting_column_w480}, 
the deformed configuration at various time instants is presented. 
Remarkably, 
the formulation adeptly captures the extremely large deformations 
encompassing the entire twisting process, 
including the recovery phase and reverse rotation, 
as anticipated.

\subsection{Electrophysiologically induced muscle contraction}
Following Refs. \cite{garcia2019new, zhang2021integrative}, 
we examine a unit cube of muscle 
characterized by an orthogonal material direction, 
where the muscle fiber and sheet directions align with the global coordinates.
The passive response is described by the Holzapfel-Ogden model, 
and the material parameters are detailed 
in Table \ref{tab:Holzapfel_Ogden_material_muscle}.
To initiate the excitation-induced response, 
a linear distribution of transmembrane potential is applied 
along the vertical direction, 
with $V_m = 0$ mV at the bottom face and $V_m = 30$ mV at the top face.
For simplicity, 
we neglect the time variation of the transmembrane potential, 
and an activation law for active stress is employed by
\begin{equation}
	T_a = -0.5 V_m.
\end{equation}
Two distinct tests involving iso- and anisotropic models 
are conducted in this study.

\begin{figure}[htb!]
	\centering
	\includegraphics[trim = 0mm 6mm 2mm 2mm, width=\textwidth]{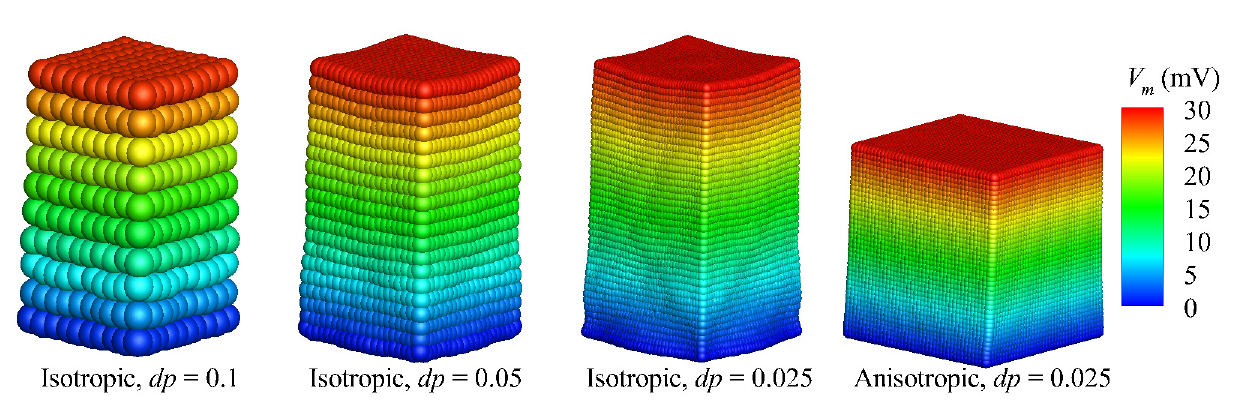}
	\caption{Muscle contraction: 
		Deformed configuration colored by transmembrane potential $V_m$ 
		obtained by the SPH-GENOG 
		when the transmembrane potential of top face $V_m = 30$ mV
		with three different spatial resolutions
		and both isotropic and anisotropic material properties.
		Note that $dp$ is the initial particle spacing.}
	\label{figs:muscle_contraction_convergence}
\end{figure}

\begin{figure}[htb!]
	\centering
	\includegraphics[trim = 0mm 10mm 2mm 2mm, width=0.5\textwidth]{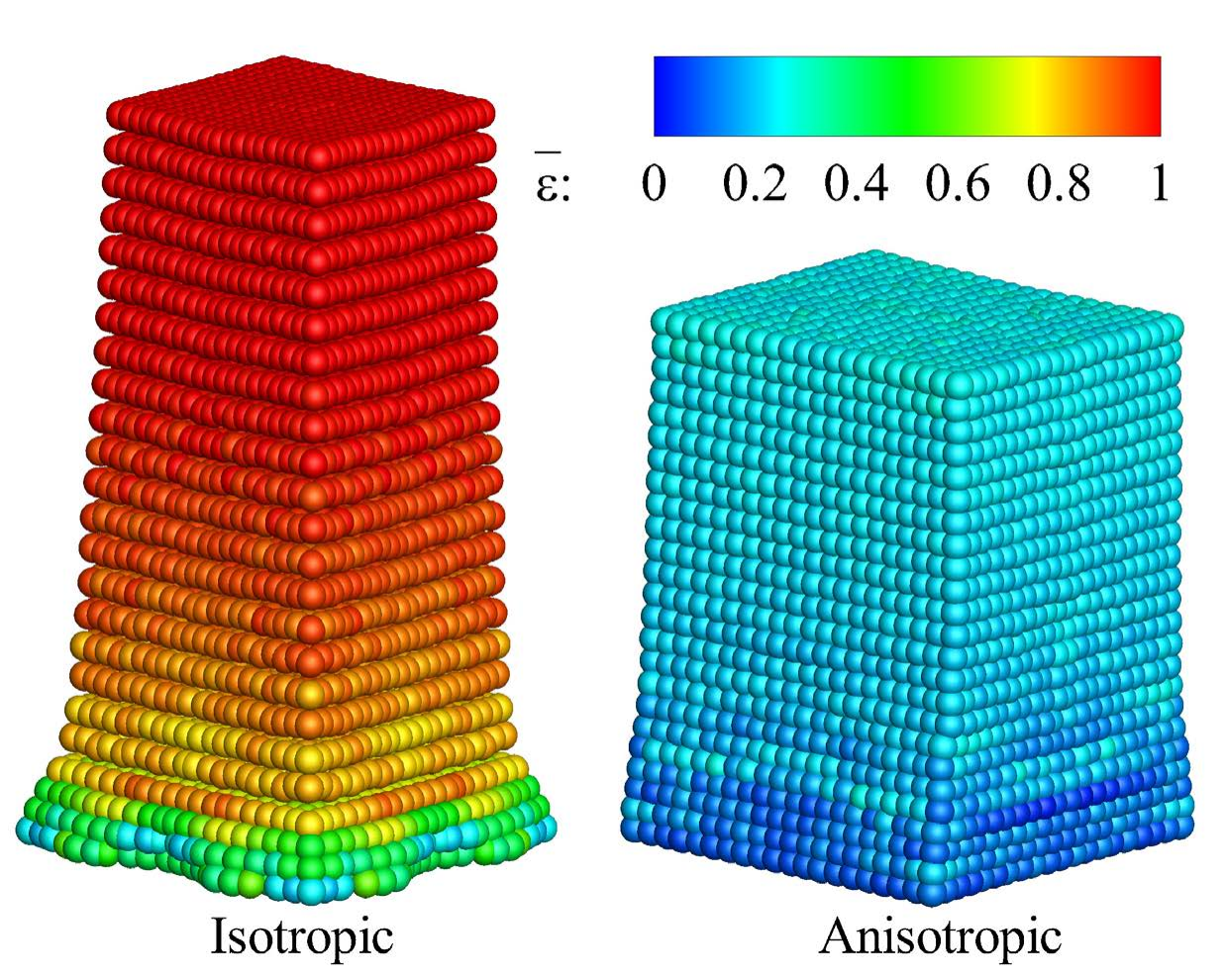}
	\caption{Muscle contraction: 
		Deformed configuration colored by von Mises strain $\bar\epsilon$ 
		obtained by the SPH-GENOG 
		when the transmembrane potential of top face is increased to $V_m = 300$ mV
		with both isotropic and anisotropic material properties.
		The spatial particle discretization is set as $dp = 0.025$.}
	\label{figs:muscle_constraction_V300}
\end{figure}

Figure \ref{figs:muscle_contraction_convergence} shows the deformed
configuration of the cubic muscle with particle refinement. 
The results showcase good convergence properties, 
and qualitative agreement is observed for the isotropic test, 
aligning well with the findings presented in Ref. \cite{garcia2019new} 
(refer to Figure 7 in their work). 
Moreover, 
Table \ref{tab:muscle_constraction_comparison} indicates 
that the displacement of the top face at fine particle resolution is 0.5355, 
demonstrating good agreement 
with the value of 0.535 reported in Ref. \cite{zhang2021integrative}.
In the case of the anisotropic test, 
deformation is reduced owing to the presence of fibers and sheets.
The transmembrane potential of top face is increased to $V_m = 300$ mV 
to further test the robustness of present formulation. 
As shown in Fig. \ref{figs:muscle_constraction_V300}, 
particle deformation and von Mises strain fields are well captured.
\begin{table}[htb!]
	\centering
	\caption{Muscle contraction: Parameters for the Holzapfel-Ogden material model. Note that the anisotropic terms are set to zero for the isotropic analysis.}
	\begin{tabular}{cccc}
		\hline
		$a = 0.059$ kPa   & $a_f = 18.472$ kPa  & $a_s = 2.841$ kPa & $a_{fs} = 0.216$ kPa \\ 
		\hline
		$b = 8.023$   & $b_f = 16.026$  & $b_s = 11.12 $ & $b_{fs} = 11.436$ \\ 
		\hline	
	\end{tabular}
	\label{tab:Holzapfel_Ogden_material_muscle}
\end{table}
\begin{table}[htb!]
	\centering
	\caption{Muscle contraction: Quantitative validation of the deformation.}
	\begin{tabular}{ccccc}
		\hline
		& $dp = 0.1$  & $dp = 0.05$ &  $dp = 0.025$ & Zhang et al. (2021) \\ 
		\hline
		Displacement & 0.4988  	  &0.5248	&0.5355 & 0.535\\
		\hline	
	\end{tabular}
	\label{tab:muscle_constraction_comparison}
\end{table}

\subsection{Taylor bar}
A 2D copper bar, 
characterized by an initial length of $L = 0.03 \, \text{m}$ 
and a height of $H = 0.006\, \text{m}$, 
is modeled for plane-strain analysis by undergoing impact against a rigid frictionless wall at time $t = 0 \, \text{s}$ with a velocity of 
$\bm{v_0} = \left(0, -227\right) ^{\operatorname{T}} \, \text{m/s}$. 
To simulate the material response, 
a hyperelastic-plastic model with linear hardening is employed. 
The material parameters include 
Young's modulus $E = 117 \, \text{GPa}$, 
density $\rho^0 = 8.930 \times 10^3 \, \text{kg/m}^3$, 
Poisson's ratio $\nu = 0.35$, 
yield stress $\tau_y = 0.4 \, \text{GPa}$, 
and hardening modulus $\kappa = 0.1 \, \text{GPa}$.
It should be noted that 
since the applied artificial damping stress $\fancy{$\tau$}_d$ 
significantly influences the deformation 
in high-velocity impact scenarios, 
$\fancy{$\tau$}_d$ used in the taylor cases is given as
\begin{equation}
	\fancy{$\tau$}_d = 0.125 \frac{\chi}{2}\frac{d \fancy{$b$}}{dt},
\end{equation}
indicating the adoption of a smaller numerical damping compared to other cases. Despite setting the CFL number to 0.1 for instability, 
which increases the computational overhead, 
the results converge rapidly even in low-resolution scenarios.

\begin{figure}[htb!]
	\centering
	\includegraphics[trim = 2mm 9mm 2mm 2mm, width=\textwidth]{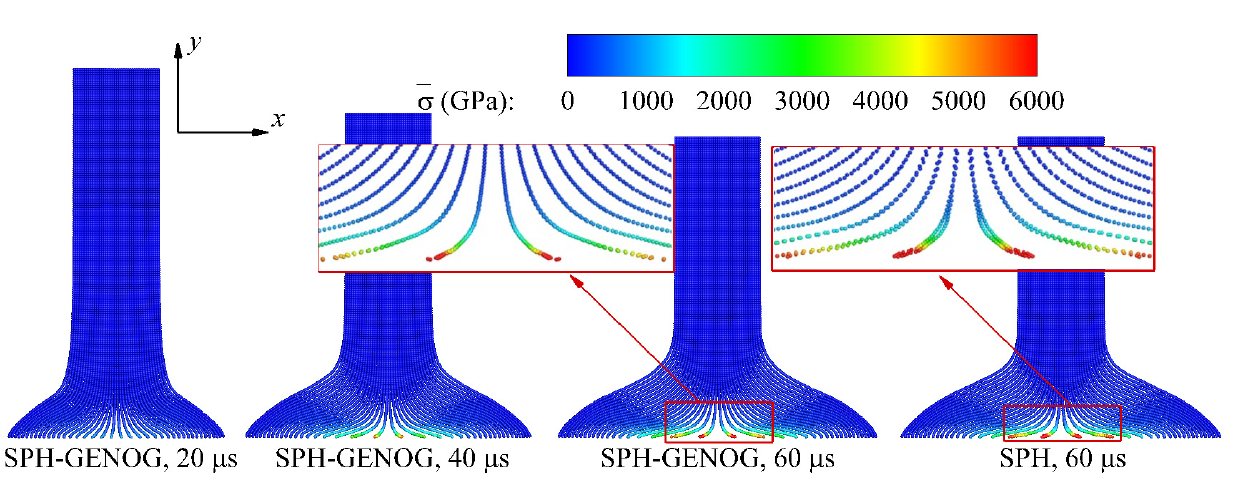}
	\caption{2D taylor bar: Deformed configuration colored by von Mises stress $\bar\sigma$ at serial temporal instants obtained by the present SPH-GENOG with initial uniform velocity $\bm{v_0} = \left(0, -227\right)^{\operatorname{T}} ~\text{m/s}$, 
		and its comparison with that of SPH. 
		The material is modeled by isotropic hardening elastic-plasticity 
		with Young's modulus $E = 117 \, \text{GPa}$, 
		density $\rho^0 = 8.930 \times 10^3 \, \text{kg/m}^3$, 
		Poisson's ratio $\nu = 0.35$, 
		yield stress $\tau_y = 0.4 \, \text{GPa}$, 
		and hardening modulus $\kappa = 0.1 \, \text{GPa}$.
		The spatial particle discretization is set as $H / dp =40$ with $H$ denoting the height of the column and $dp$ the initial particle spacing.}
	\label{figs:2D_taylor_bar_deformation_comparison}
\end{figure}
\begin{figure}[htb!]
	\centering
	\includegraphics[trim = 0mm 6mm 2mm 2mm, width=\textwidth]{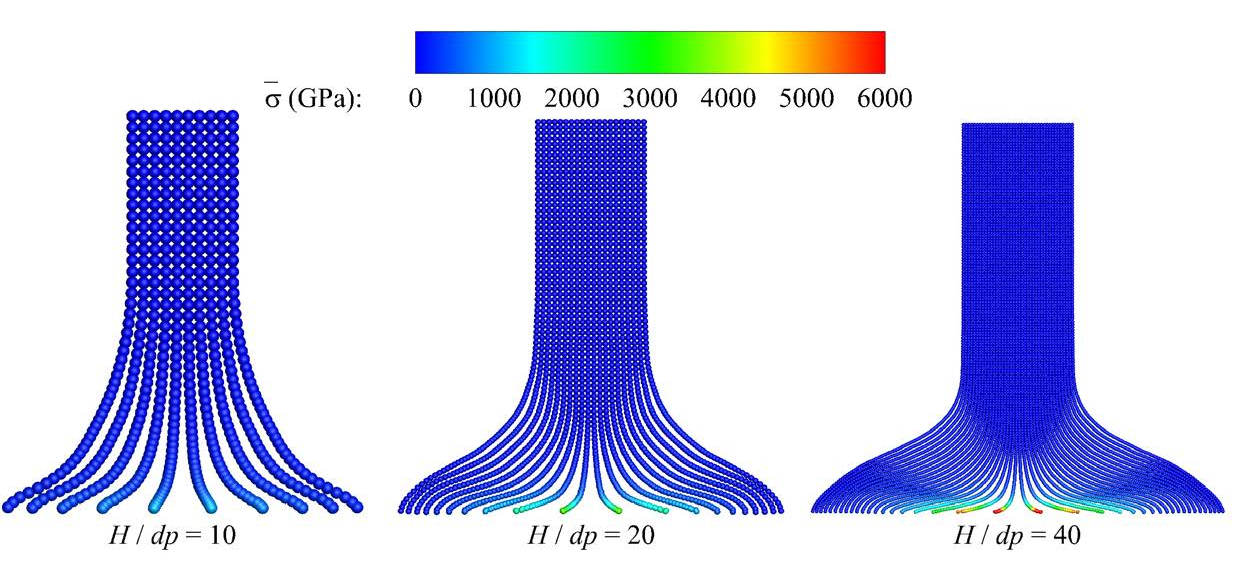}
	\caption{2D taylor bar: 
		Deformed configuration colored by von Mises stress $\bar\sigma$ 
		obtained by the SPH-GENOG 
		with three different spatial resolutions 
		and the initial uniform velocity $\bm{v_0} = \left(0, -227\right) ^{\operatorname{T}} ~\text{m/s}$.}
	\label{figs:2D_taylor_bar_convergence}
\end{figure}
\begin{figure}[htb!]
	\centering
	\includegraphics[trim = 0mm 6mm 2mm 2mm, width=\textwidth]{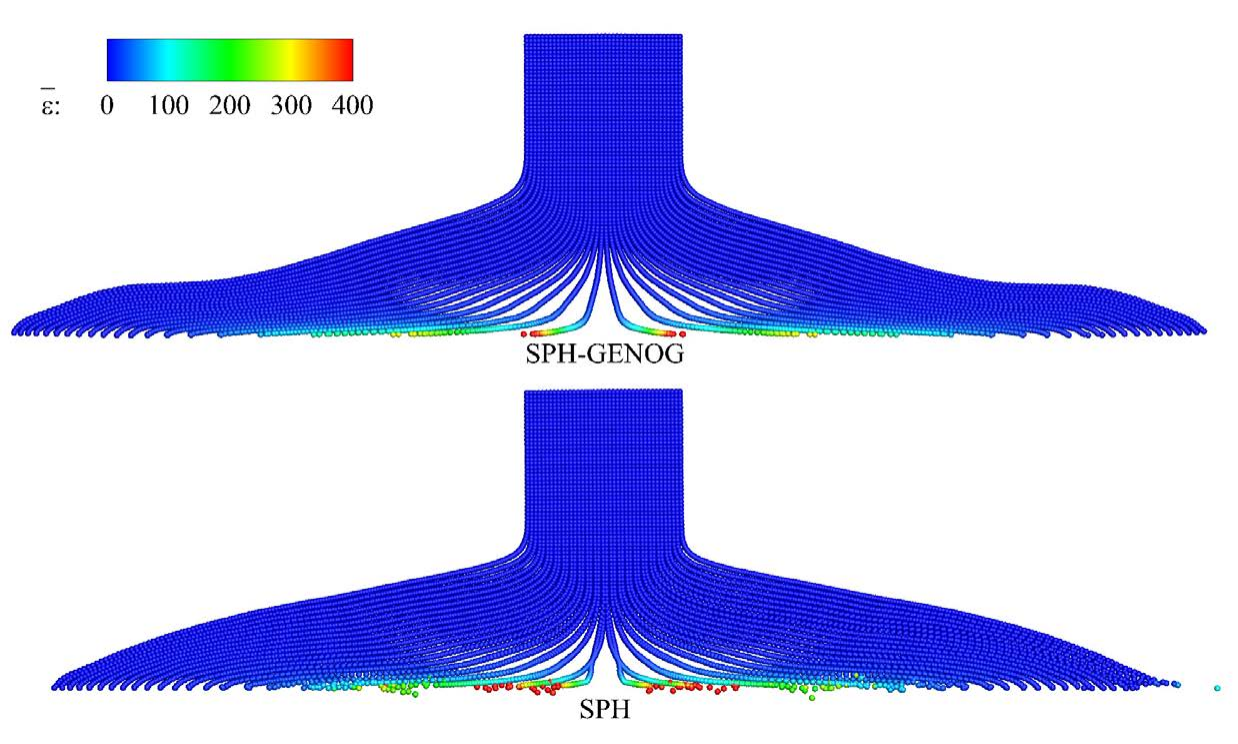}
	\caption{2D taylor bar: 
		Deformed configuration colored by von Mises strain $\bar\epsilon$ 
		obtained by the SPH-GENOG (top panel) and SPH (bottom panel)
		with initial uniform velocity $\bm{v_0} = \left(0, -400\right) ^{\operatorname{T}} ~\text{m/s}$. 
	The spatial particle discretization is set as $H / dp =40$.}
	\label{figs:2D_taylor_bar_v400}
\end{figure}
Figure \ref{figs:2D_taylor_bar_deformation_comparison} shows 
the deformed configuration of 2D taylor bar
at different time instants with von Mises stress contour 
obtained by the SPH-GENOG, 
and its comparison with that simulated by SPH 
when the time $t = 60\,\mu\text{s}$.
While both simulations exhibit satisfactory performance
and produce comparable results in terms of deformation and stress patterns,  
SPH-GENOG demonstrates a more uniform particle distribution compared to SPH.
A convergence study is undertaken, 
incrementally refining the spatial resolution 
from $H/dp = 10$ to $H/dp = 20$ and $H/dp = 40$.
The convergence properties of both deformation and von Mises stress  $\bar\sigma$
are shown in Fig. \ref{figs:2D_taylor_bar_convergence}.
A significantly more challenging problem is studied 
by increasing the initial velocity 
to $\bm{v_0} = \left(0, -400\right) ^{\operatorname{T}} ~\text{m/s}$.
As illustrated in Fig. \ref{figs:2D_taylor_bar_v400}, 
the unstabilized results from SPH exhibit noticeable particle disorder. 
Conversely, 
the outcomes obtained through SPH-GENOG demonstrate 
an orderly particle distribution and a smooth strain field, 
even in the presence of significant strain 
(the maximum von Mises strain exceeds 400). 

\begin{figure}[htb!]
	\centering
	\includegraphics[trim = 0mm 6mm 2mm 4mm]{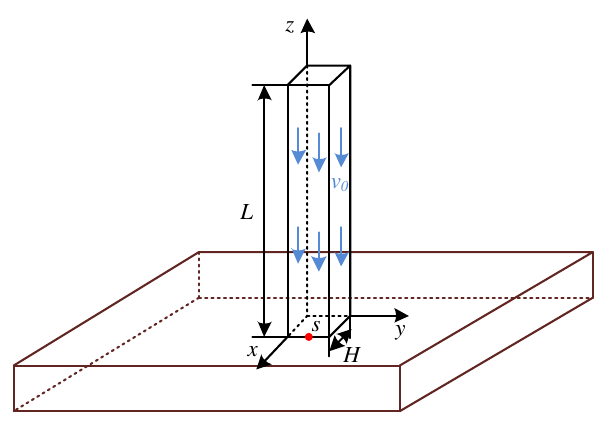}
	\caption{3D taylor bar: Problem setup.}
	\label{figs:3D_taylor_bar_setup}
\end{figure}
\begin{figure}[htb!]
	\centering
	\includegraphics[trim = 2mm 4mm 2mm 2mm, width=\textwidth]{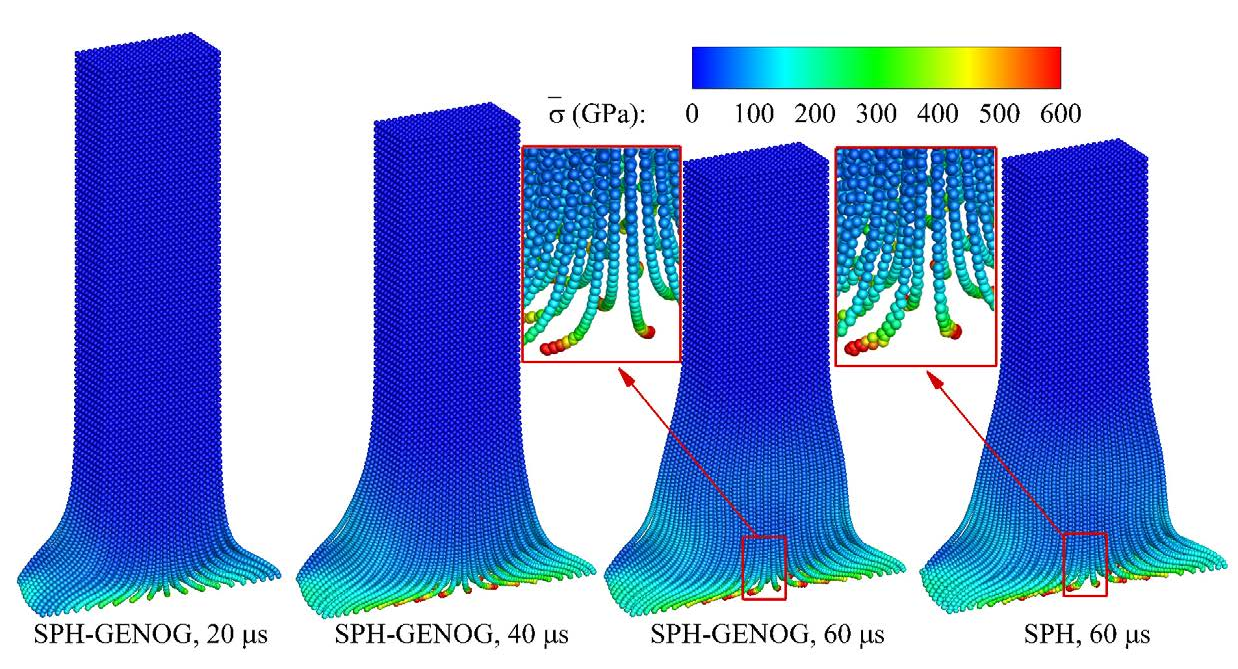}
	\caption{3D taylor bar: Deformed configuration colored by von Mises stress $\bar\sigma$ at serial temporal instants obtained by the present SPH-GENOG with initial uniform velocity $\bm{v_0} = \left(0, 0, -227\right)^{\operatorname{T}} ~\text{m/s}$, 
		and its comparison with that of SPH. 
		The material is modeled by isotropic hardening elastic-plasticity 
		with Young's modulus $E = 117 \, \text{GPa}$, 
		density $\rho^0 = 8.930 \times 10^3 \, \text{kg/m}^3$, 
		Poisson's ratio $\nu = 0.35$, 
		yield stress $\tau_y = 0.4 \, \text{GPa}$, 
		and hardening modulus $\kappa = 0.1 \, \text{GPa}$.
		The spatial particle discretization is set as $H / dp =20$ with $H$ denoting the height of the column and $dp$ the initial particle spacing.}
	\label{figs:3D_taylor_bar_deformation_comparison}
\end{figure}
\begin{figure}[htb!]
	\centering
	\includegraphics[trim = 0mm 6mm 2mm 2mm, width=\textwidth]{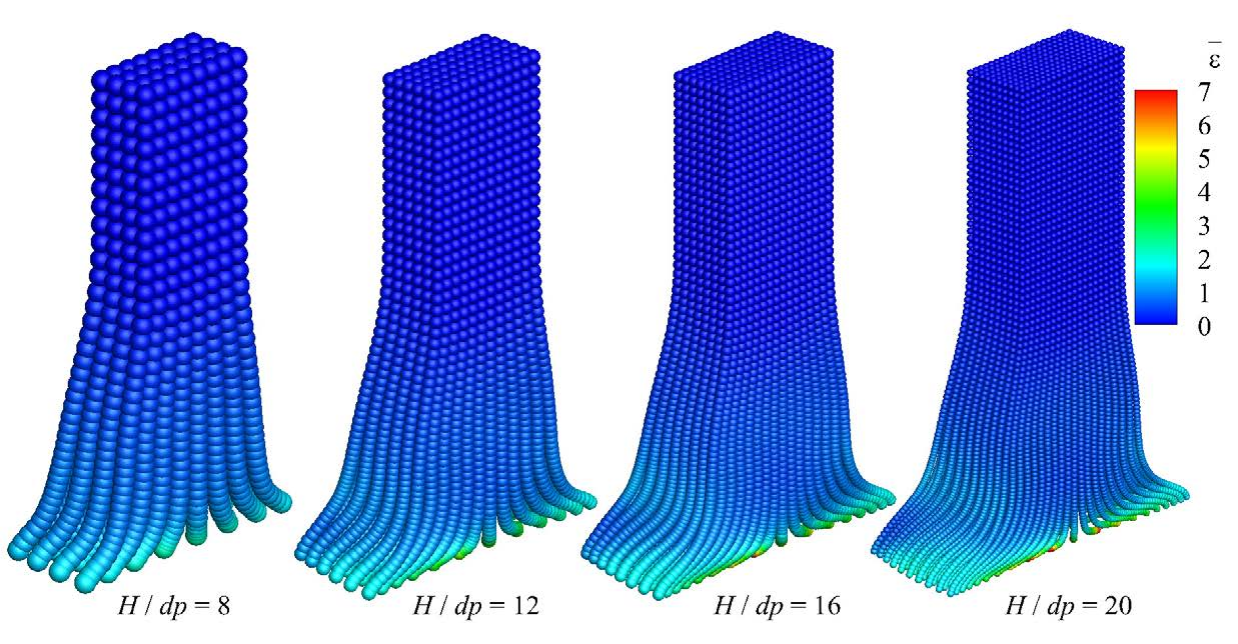}
	\caption{3D taylor bar: 
		Deformed configuration colored by von Mises strain $\bar\epsilon$ 
		obtained by the SPH-GENOG 
		with four different spatial resolutions 
		and the initial uniform velocity $\bm{v_0} = \left(0, 0, -227\right) ^{\operatorname{T}} ~\text{m/s}$.}
	\label{figs:3D_taylor_bar_convergence_contour}
\end{figure}
\begin{figure}[htb!]
	\centering
	\includegraphics[trim = 2mm 6mm 2mm 2mm, width=0.5 \textwidth] {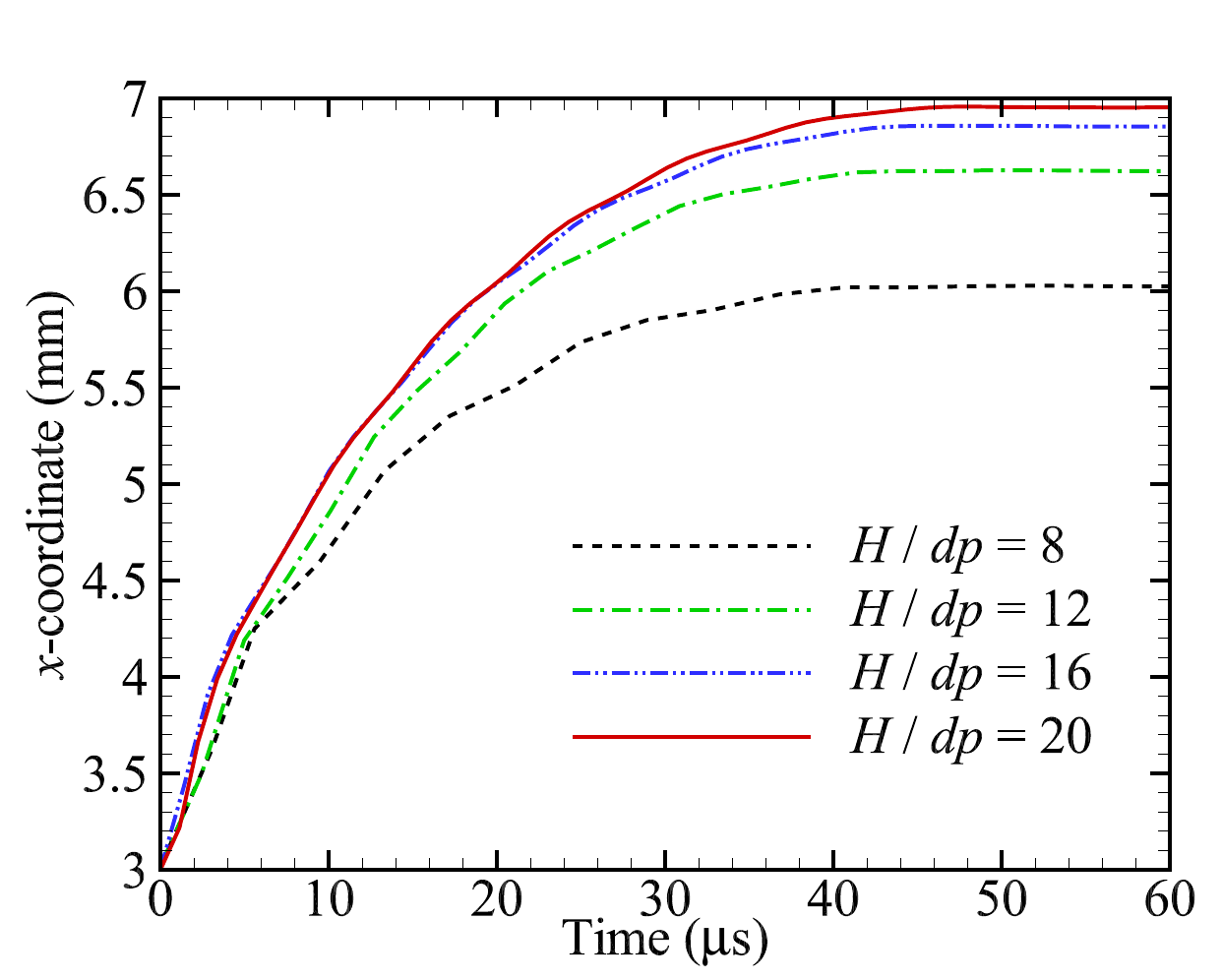}
	\caption{3D taylor bar: Time history of the horizontal position $x$ observed at node $S$ obtained by the SPH-GENOG with initial uniform velocity $\bm{v_0} = \left(0, 0, -227\right)^{\operatorname{T}} ~\text{m/s}$ under four different resolutions.}
	\label{figs:3D_taylor_bar_convergence_curves}
\end{figure}
\begin{figure}[htb!]
	\centering
	\includegraphics[trim = 0mm 6mm 2mm 2mm, width=\textwidth]{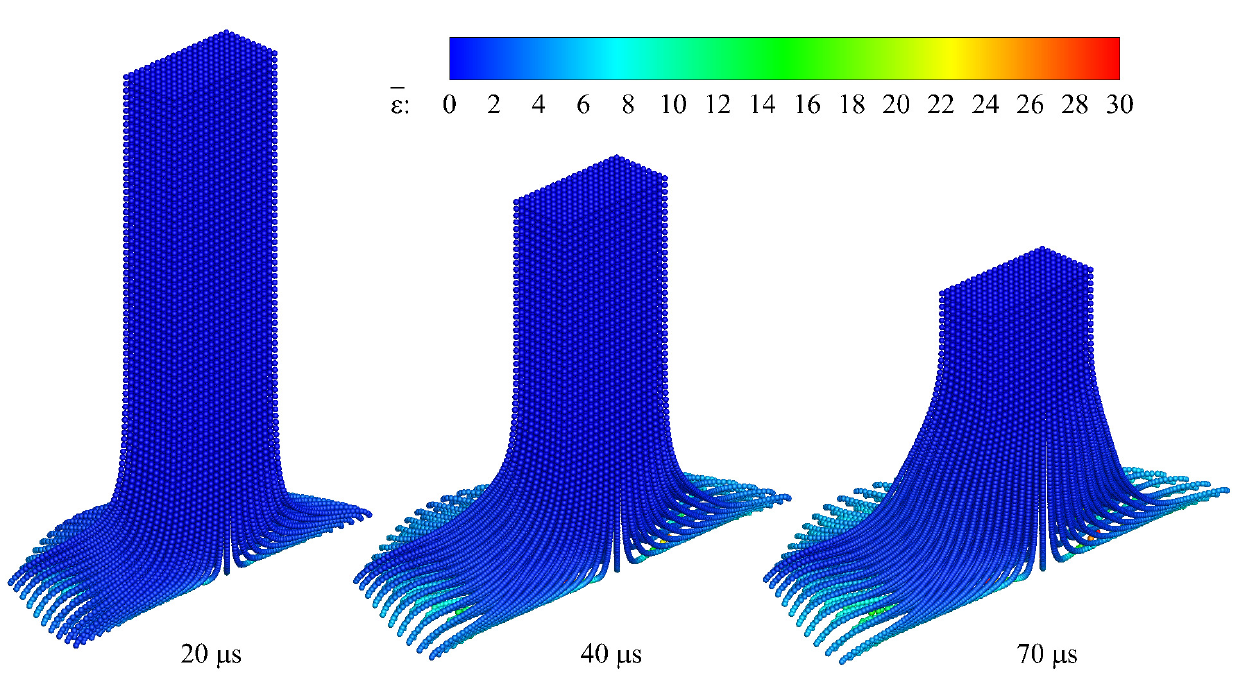}
	\caption{3D taylor bar: 
		Deformed configuration colored by von Mises strain $\bar\epsilon$ 
		at serial temporal instants
		obtained by the SPH-GENOG with initial uniform velocity $\bm{v_0} = \left(0, 0, -350\right) ^{\operatorname{T}} ~\text{m/s}$. 
		The spatial particle discretization is set as $H / dp =20$.}
	\label{figs:3D_taylor_bar_v350}
\end{figure}
The 2D taylor bar is expanded to a 3D analysis, 
featuring a squared cross-section with dimensions of $0.006 \times 0.006 \, \text{m}$, 
as shown in Fig. \ref{figs:3D_taylor_bar_setup}. 
Figure \ref{figs:3D_taylor_bar_deformation_comparison} 
illustrates the deformed configuration of the 3D taylor bar at various time instants, 
accompanied by von Mises stress contours obtained through SPH-GENOG, 
and a comparative analysis is presented against the simulation 
conducted by SPH at $t = 60\,\mu\text{s}$.
Although both simulations demonstrate good performance 
and yield comparable results in terms of deformation and stress patterns, SPH-GENOG still exhibits a more uniform particle distribution compared to SPH.
A sequence of particle refinement analyzes, 
from $H/dp = 8$ to $H/dp = 12$, $H/dp = 16$ and $H/dp = 20$, 
is also conducted. 
As presented in Fig. \ref{figs:3D_taylor_bar_convergence_contour}, 
the good convergence characteristics of both deformation 
and the von Mises stress $\bar\sigma$ are observed. 
For further convergence analysis and quantitative validation, 
Fig. \ref{figs:3D_taylor_bar_convergence_curves} illustrates the temporal evolution of the $x$-axis position of point $S$ marked in Fig. \ref{figs:3D_taylor_bar_setup}.
It is evident from observation that the displacement converges rapidly,
approximating a second-order rate, 
and the $x$-axis position of the highest resolution is $x = 6.953\, \text{mm}$,
aligning closely with the results in Ref. \cite{haider2017first}.
A more demanding scenario is investigated
by increasing the initial velocity 
to $\bm{v_0} = \left(0, 0, -350\right) ^{\operatorname{T}} ~\text{m/s}$.
As depicted in Fig. \ref{figs:3D_taylor_bar_v350}, 
the results obtained through SPH-GENOG 
still exhibit an organized particle distribution and a smooth stress field.

\begin{figure}[htb!]
	\centering
	\includegraphics[trim = 0mm 6mm 2mm 2mm, width=\textwidth]{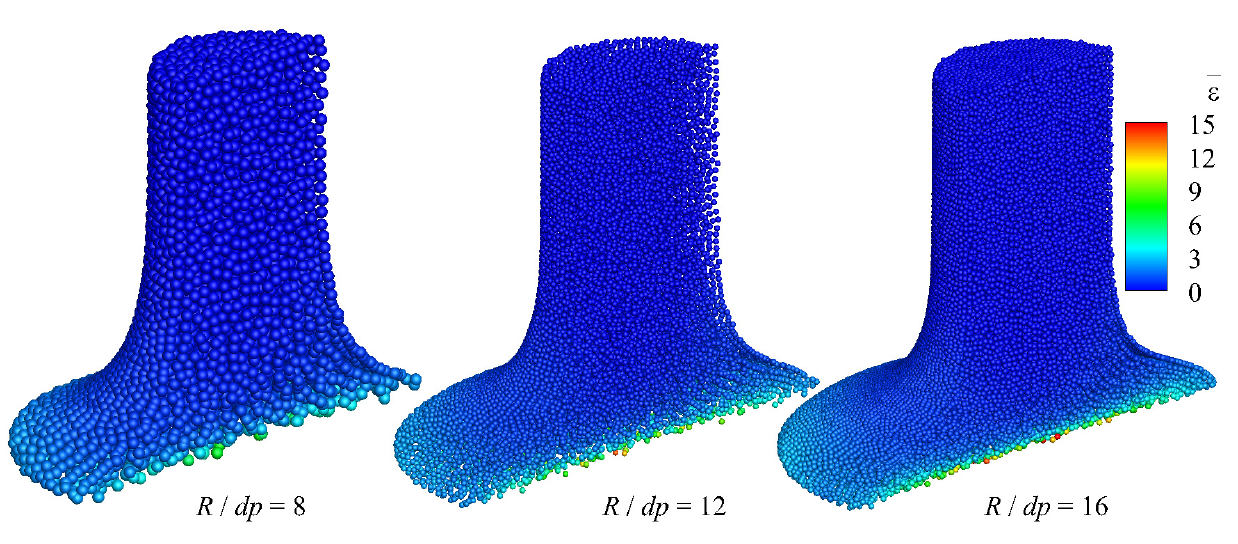}
	\caption{3D round taylor bar: 
		Deformed configuration colored by von Mises strain $\bar\epsilon$ 
		obtained by the SPH-GENOG 
		with three different spatial resolutions 
		and the initial uniform velocity $\bm{v_0} = \left(0, 0, -373\right) ^{\operatorname{T}} ~\text{m/s}$.
		The perfect plastic material is modeled with density $\rho_0 = 2700 \, \text{kg/m}^3$, 
		Young's modulus $E = 78.2 \, \text{GPa}$, Poisson's ratio $\nu = 0.3$, 
		and yield stress $\tau_y = 0.29 \, \text{GPa}$.
		Note that $R$ is the radius of bar and $dp$ the initial particle spacing.}
	\label{figs:3D_round_bar_convergence}
\end{figure}
Following Refs. \cite{taylor1948use, chen1996reproducing}, 
we now investigate a round aluminum bar with the initial length $L = 2.346 \, \text{cm}$ and radius $R = 0.391 \, \text{cm}$. 
The material is modeled by perfect plasticity, 
i.e., hardening modulus $\kappa = 0 \, \text{Pa}$, 
with initial density $\rho_0 = 2700 \, \text{kg/m}^3$, 
Young's modulus $E = 78.2 \, \text{GPa}$, Poisson's ratio $\nu = 0.3$, 
and yield stress $\tau_y = 0.29 \, \text{GPa}$. 
The initial impact velocity is set as $\bm{v_0} = \left(0, 0, -373\right) ^{\operatorname{T}} ~\text{m/s}$. 
A convergence study is conducted with three resolutions, 
$R/dp = 8$, $R/dp = 12$ and $R/dp = 16$.
As shown in Fig. \ref{figs:3D_round_bar_convergence}, 
the good convergence characteristics of both deformation 
and the von Mises strain $\bar\epsilon$ are observed. 
For quantitative validation, 
Table \ref{tab:3D_round_bar_comparison} summarizes the deformation under various resolutions 
and compares it with the results from Ref. \cite{chen1996reproducing}. 
Favorable convergence properties and high accuracy are observed.
\begin{table}[htb!]
	\centering
	\caption{3D round taylor bar: Quantitative validation of deformed geometries for perfect plastic material.}
	\begin{tabular}{ccccc}
		\hline
		& $R / dp = 8$  & $R / dp = 12$ & $R / dp = 16$ & Chen et al. (1996) \\ 
		\hline
		Length (cm)	& 1.4908  	  & 1.4631	&1.4546	&1.454\\
		Radius (cm)	& 0.9075  	  & 0.9323	&0.9616	&1.051\\
		\hline	
	\end{tabular}
	\label{tab:3D_round_bar_comparison}
\end{table}

\subsection{Necking bar}
\begin{figure}[htb!]
	\centering
	\includegraphics[trim = 0mm 6mm 2mm 0mm]{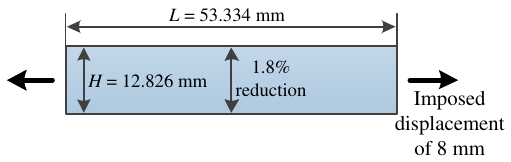}
	\caption{Necking bar: Problem setup.}
	\label{figs:Necking_bar_setup}
\end{figure}

In this section, 
we examine a plane-strain bar undergoing uniform extension, 
a standard test problem analyzed in Refs. \cite{simo2006computational, taylor2011isogeometric, elguedj2014isogeometric}.
The bar dimensions are length $L = 53.334\, \text{mm}$ 
and height $H = 12.826\, \text{mm}$. 
To control the location of the necking, 
the center dimension of the bar is reduced 
to 0.982 of the side height (1.8\% reduction), 
as shown in Fig. \ref{figs:Necking_bar_setup}.
A total displacement of 8 mm is applied on the constrained boundary particles,
an additional 4 layers of particles on both sides. 
The bar exhibits elastic deformation governed by the Neo-Hookean law 
and the plastic response characterized by the nonlinear isotropic hardening law. Material parameters are detailed in Table \ref{tab:stretching_parameters}.
\begin{table}[htb]
	\centering
	\caption{Necking bar: Non-linear hardening elastic-plastic material parameters.}
	\begin{tabular}{cc} 
		\hline		
		Parameters & Value  \\  
		\hline	
		Shear modulus   & 80.1938 GPa\\  
		Bulk modulus  & 164.21  GPa  \\ 
		Initial flow stress 	&450 MPa \\ 
		Saturation flow stress	&715 MPa\\ 
		Saturation exponent 	& 16.93 \\   
		Linear hardening coefficient & 129.24 MPa\\ 
		\hline	
	\end{tabular}
	\label{tab:stretching_parameters}
\end{table}
\begin{figure}[htb!]
	\centering
	\includegraphics[trim = 2mm 4mm 2mm 2mm, width=0.78 \textwidth]{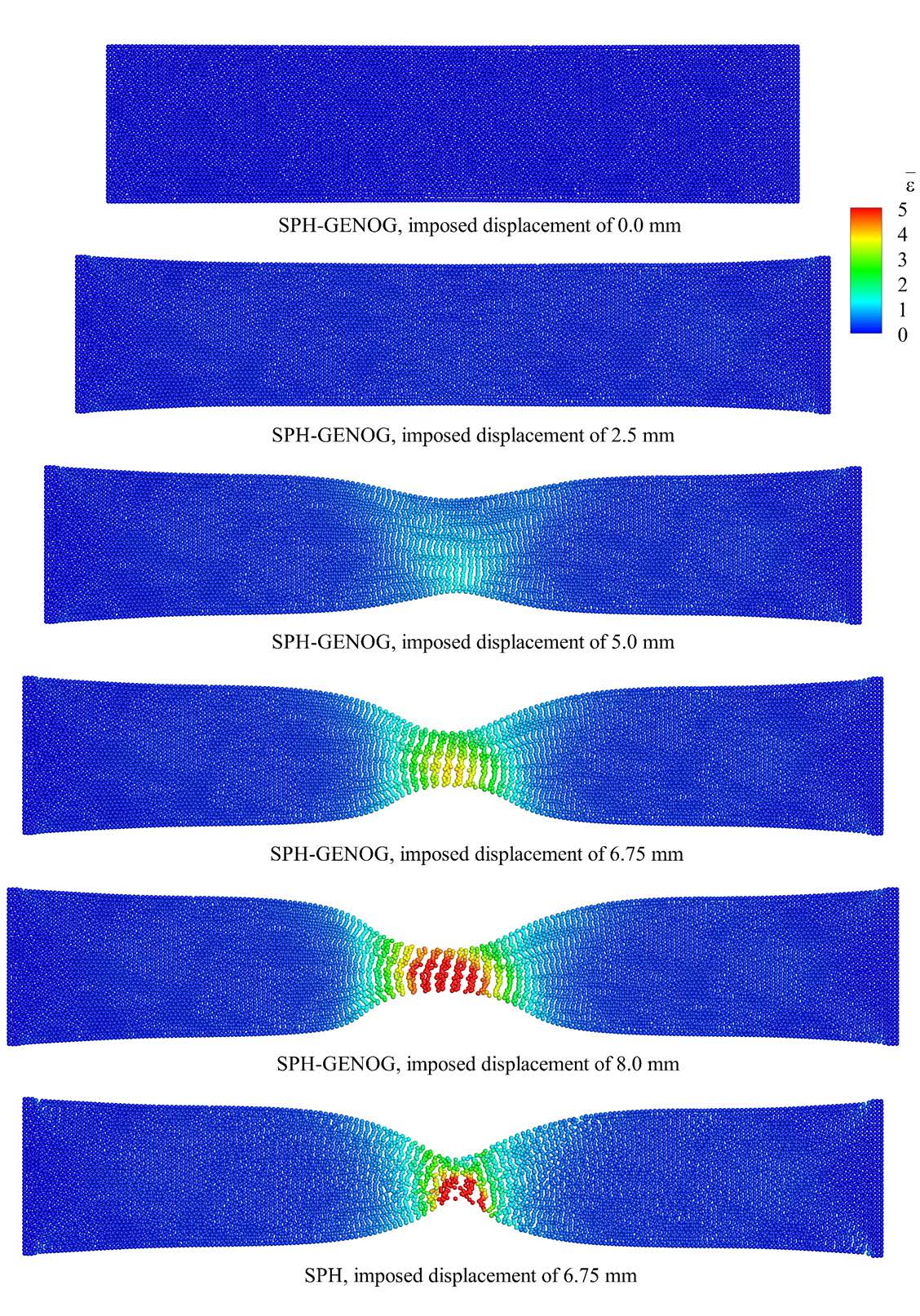}
	\caption{Necking bar: Deformed configuration colored by von Mises strain $\bar\epsilon$ at various instants obtained by the present SPH-GENOG, 
		and its comparison with that of SPH. 
		The spatial particle discretization is set as $H / dp =60$ with $H$ denoting the height of bar and $dp$ the initial particle spacing.}
	\label{figs:2D_necking_stress}
\end{figure}
\begin{figure}[htb!]
	\centering
	\includegraphics[trim = 0mm 6mm 2mm 2mm, width=\textwidth]{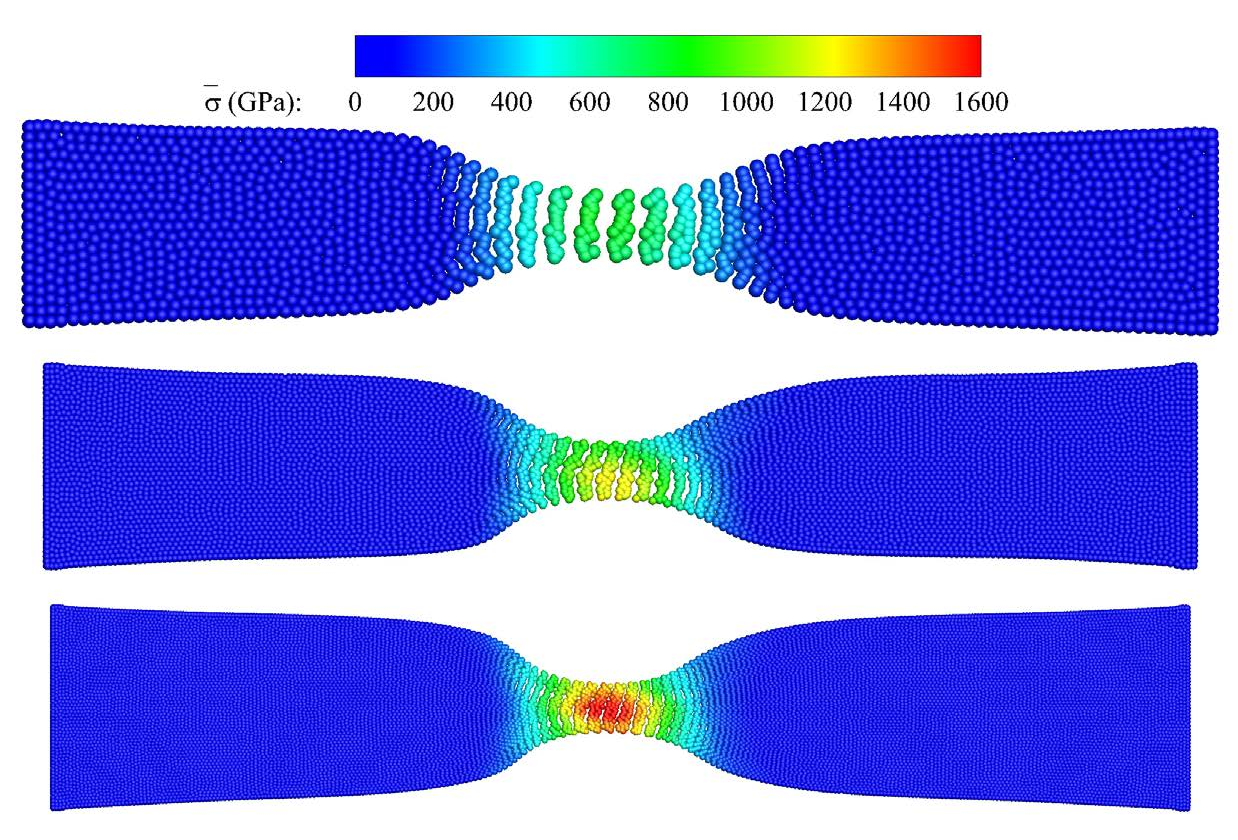}
	\caption{Necking bar: A sequence of particle refinement analyzes
		using the present SPH-GENOG. 
		Three different spatial resolutions,
		$H / dp =20$, $H / dp =40$ and $H / dp =60$, are applied.}
	\label{figs:2D_necking_stress_convergence}
\end{figure}
\begin{figure}[htb!]
	\centering
	\includegraphics[trim = 2mm 6mm 2mm 2mm, width=0.5\textwidth] {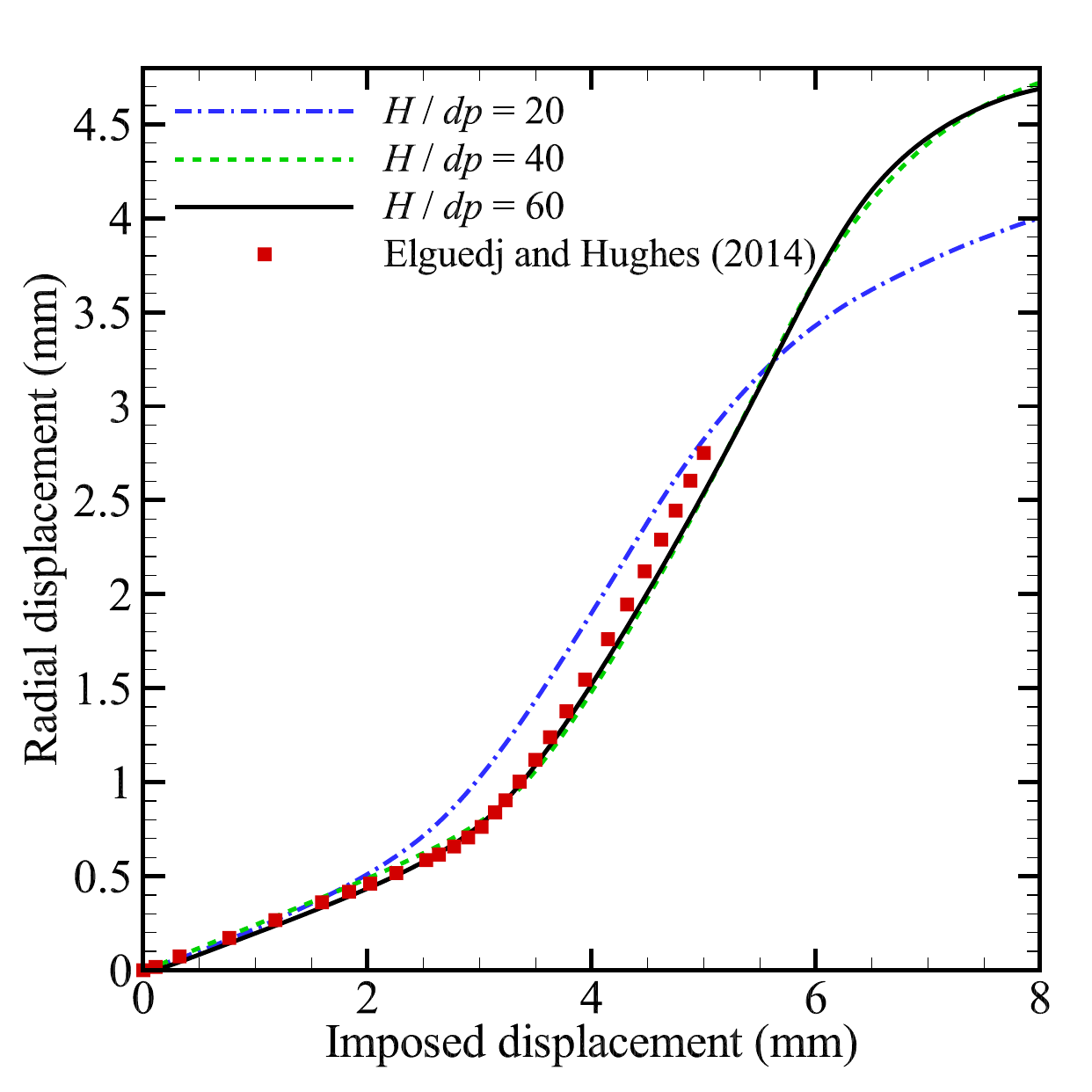}
	\caption{Necking bar: Necking displacement versus imposed displacement obtained by the present SPH-GENOG with three different spatial resolutions, and its comparison with that of Elguedj and Hughes \cite{elguedj2014isogeometric}.}
	\label{figs:2D_necking_convergence}
\end{figure}
\begin{figure}[htb!]
	\centering
	\includegraphics[trim = 2mm 6mm 2mm 2mm, width=0.5\textwidth] {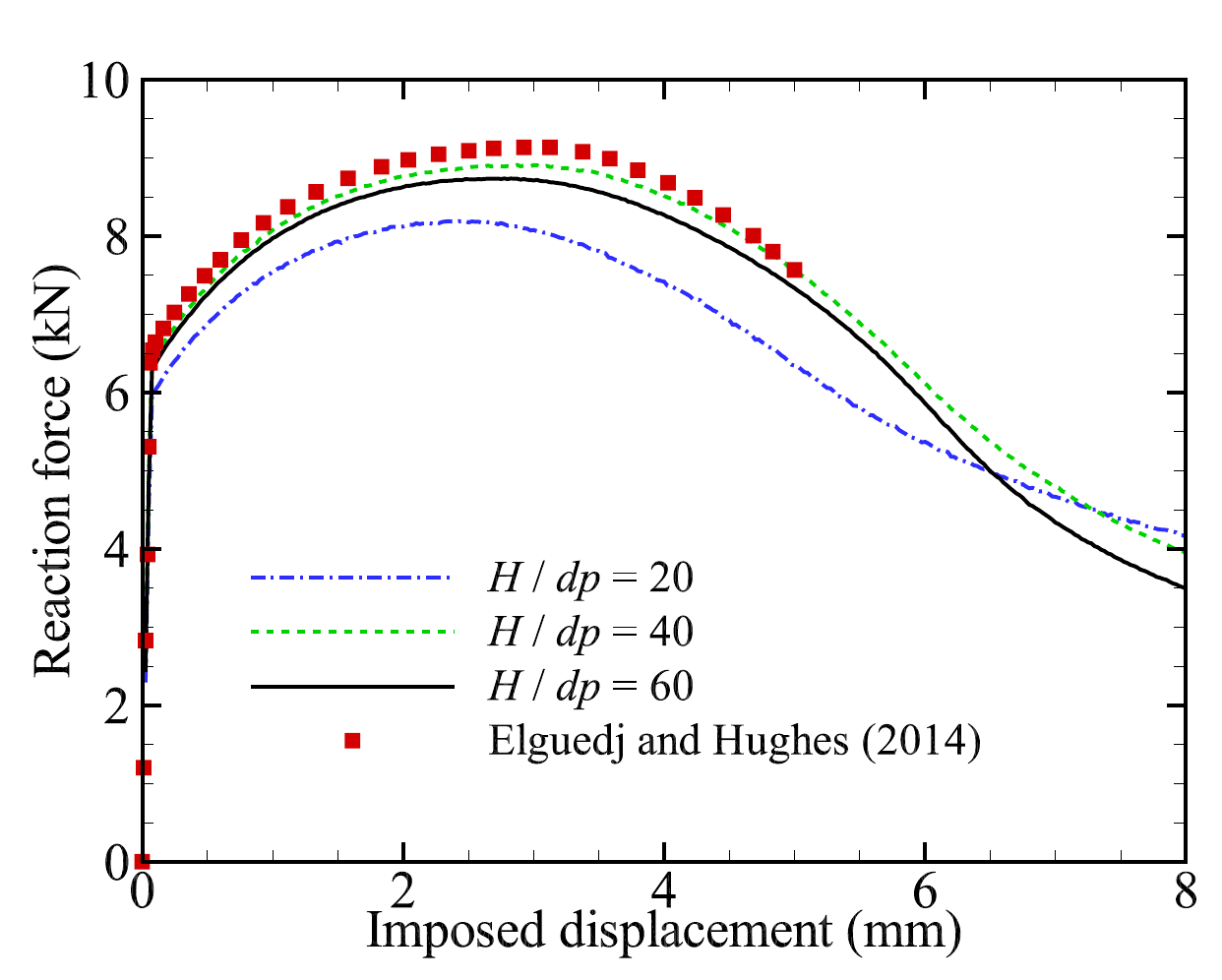}
	\caption{Necking bar: Reaction force versus imposed displacement obtained by the present SPH-GENOG with three different spatial resolutions, and its comparison with that of Elguedj and Hughes \cite{elguedj2014isogeometric}.}
	\label{figs:2D_necking_convergence-2}
\end{figure}
\begin{figure}[htb!]
	\centering
	\includegraphics[trim = 2mm 6mm 2mm 2mm, width=0.5\textwidth] {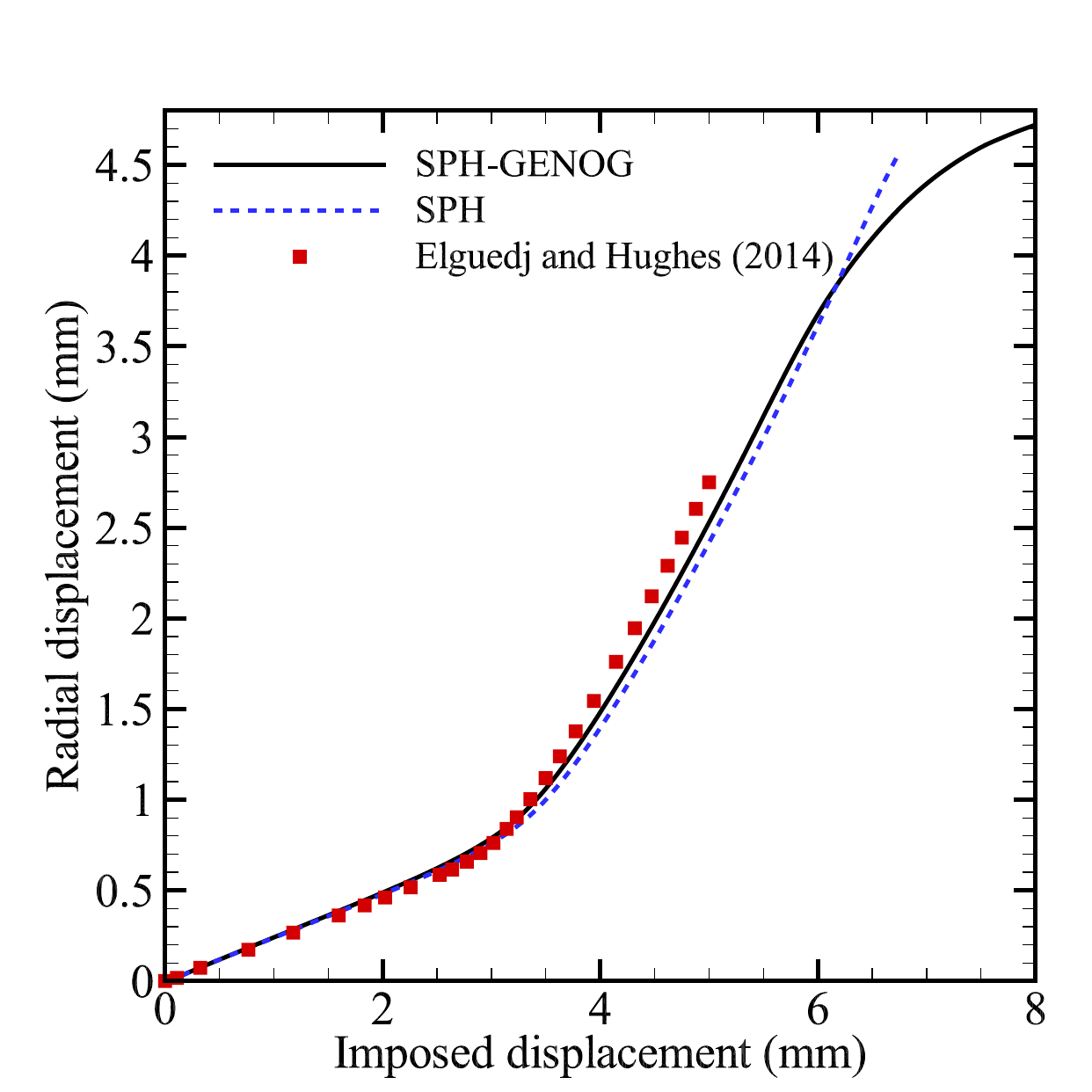}
	\caption{Necking bar: Necking displacement versus imposed displacement obtained by the SPH-GENOG and SPH, and their comparison with that of Elguedj and Hughes \cite{elguedj2014isogeometric}.
		The spatial particle discretization is $H / dp = 40$.}
	\label{figs:2D_necking_Comparison_SPH-HC_SPH}
\end{figure}
\begin{figure}[htb!]
	\centering
	\includegraphics[trim = 2mm 6mm 2mm 2mm, width=0.5\textwidth] {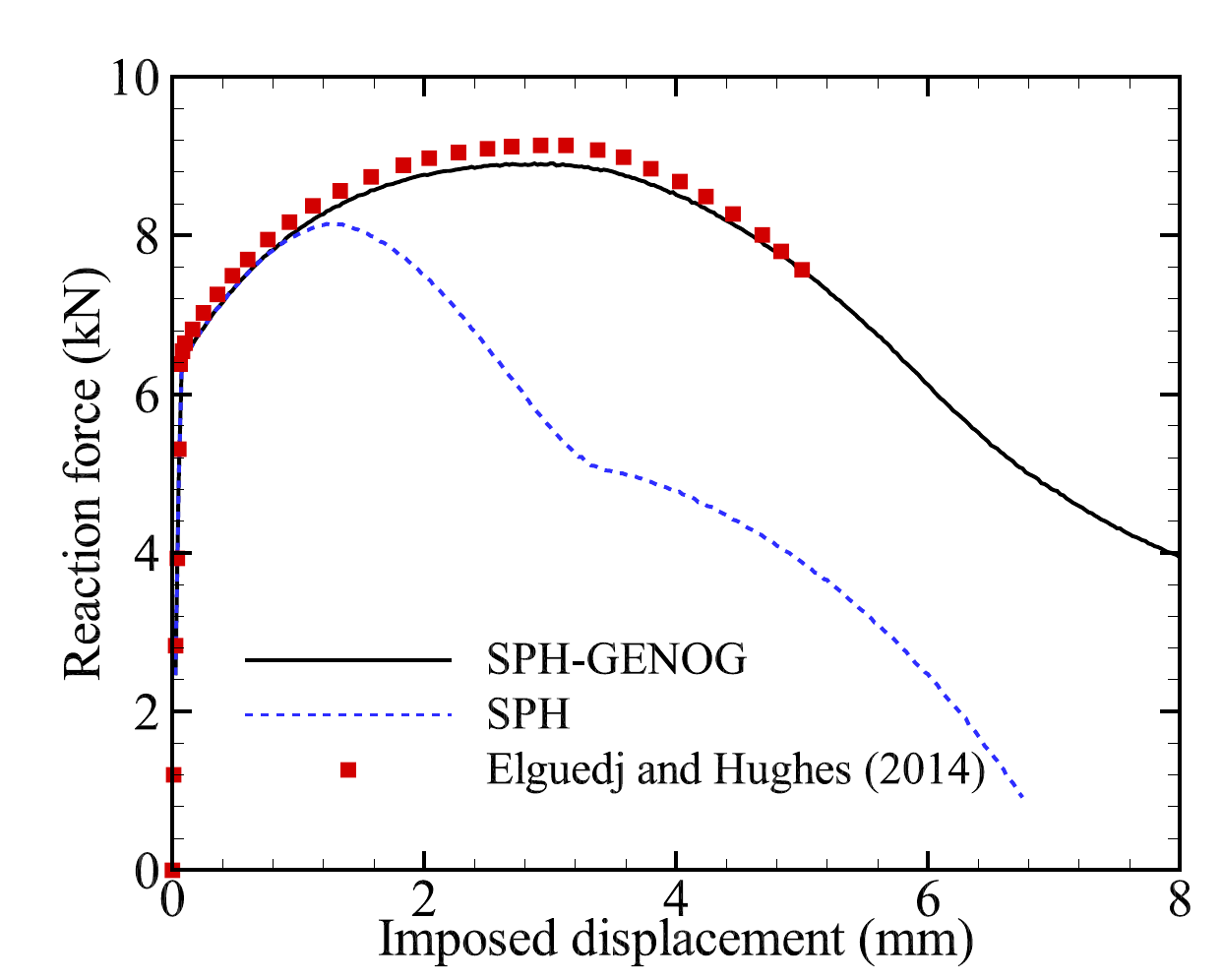}
	\caption{Necking bar: Reaction force versus imposed displacement obtained by the SPH-GENOG and SPH, and their comparison with that of Elguedj and Hughes \cite{elguedj2014isogeometric}.
		The spatial particle discretization is $H / dp = 40$.}
	\label{figs:2D_necking_Comparison_SPH-HC_SPH_2}
\end{figure}
Figure \ref{figs:2D_necking_stress} depicts the deformed configuration 
of the necking bar at different instants, 
featuring von Mises strain contours, 
obtained through SPH-GENOG, 
and a comparative analysis with the simulation performed by SPH 
under the applied displacement of 6.75 mm. 
While SPH exhibits noticeable particle disorder, 
SPH-GENOG presents commendable performance in capturing deformation 
and strain patterns with a organized particle distribution. 
It should be noted that, 
despite the symmetry of this necking bar problem, 
the strain field exhibits asymmetry 
owing to the initial asymmetric particle distribution.
A series of particle refinement analyses are performed, 
with the spatial resolution varying from 
$H/dp = 20$ to $H/dp = 40$ and $H/dp = 60$. 
The results, depicted in Fig. \ref{figs:2D_necking_stress_convergence}, 
reveal quite good convergence properties in both deformation and von Mises stress $\bar\sigma$, 
reinforcing the reliability of the simulation outcomes. 
For a more comprehensive convergence analysis and quantitative validation, 
Figs. \ref{figs:2D_necking_convergence} and \ref{figs:2D_necking_convergence-2} present 
the necking displacement of the bar center dimension 
and the corresponding reaction force 
exerted by the material in response to the applied tensile load. 
These results are compared with the highest-order finite element outcomes 
reported in Ref. \cite{elguedj2014isogeometric}. 
The significant convergence properties and accuracy are noted, 
reaffirming the reliability of the simulation.
And we can observe that
after a short elastic response, 
indicated by the initial linear segment of the reaction force curve, 
the bar transitions to plastic deformation, 
marked by a slowly increasing reaction force. 
Subsequently, 
the deformation shifts to a mode 
where plastic effects concentrate in the necking area, 
leading to a decrease in the reaction force.
Figs. \ref{figs:2D_necking_Comparison_SPH-HC_SPH} and \ref{figs:2D_necking_Comparison_SPH-HC_SPH_2} 
present a comparative analysis of displacement and reaction force curves 
between SPH-GENOG and SPH. 
Notably, 
SPH exhibits inaccuracies after a stretching of 1.2 mm due to hourglass modes,
while SPH-GENOG maintains accurate performance throughout.

\subsection{Oobleck octopus}
\begin{figure}[htb!]
	\centering
	\includegraphics[trim = 0mm 6mm 2mm 2mm, width=0.5\textwidth]{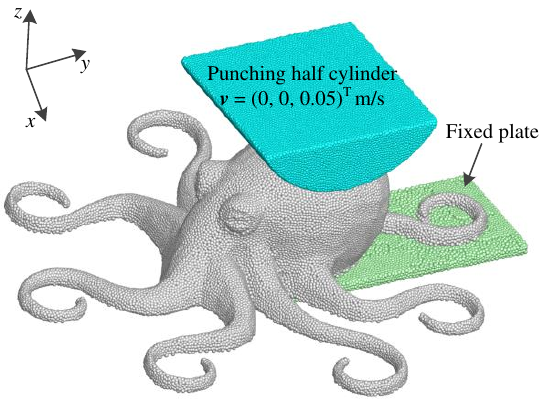}
	\caption{Oobleck octopus: Initial configuration.}
	\label{figs:octopus_0s}
\end{figure}

In this section, 
we analyze mechanical behaviors of an octopus 
made of shear thickening oobleck, 
a viscoplastic material. 
The octopus undergoes deformation 
under its own gravity and the punch from a half-cylinder, 
as illustrated in Fig. \ref{figs:octopus_0s}. 
Note that the half-cylinder stops punching after 0.3 seconds.
The material properties of oobleck are provided 
in Table \ref{tab:oobleck_parameters} \cite{yue2015continuum}.
\begin{table}[htb]
	\centering
	\caption{Oobleck octopus: Viscoplastic material parameters.}
	\begin{tabular}{cc} 
		\hline		
		Parameters & Value  \\  
		\hline	
		Density & 1000.0 $\text{kg/m}^3$\\
		Shear modulus   & 11.2 kPa\\  
		Bulk modulus  & 109.0  kPa  \\ 
		Yield stress 	& 0.1 Pa \\ 
		Viscosity	& 10 \\ 
		Herschel Bulkley power	& 2.8\\   
		\hline	
	\end{tabular}
	\label{tab:oobleck_parameters}
\end{table}
\begin{figure}[htb!]
	\centering
	\includegraphics[trim = 0mm 6mm 2mm 2mm, width=\textwidth]{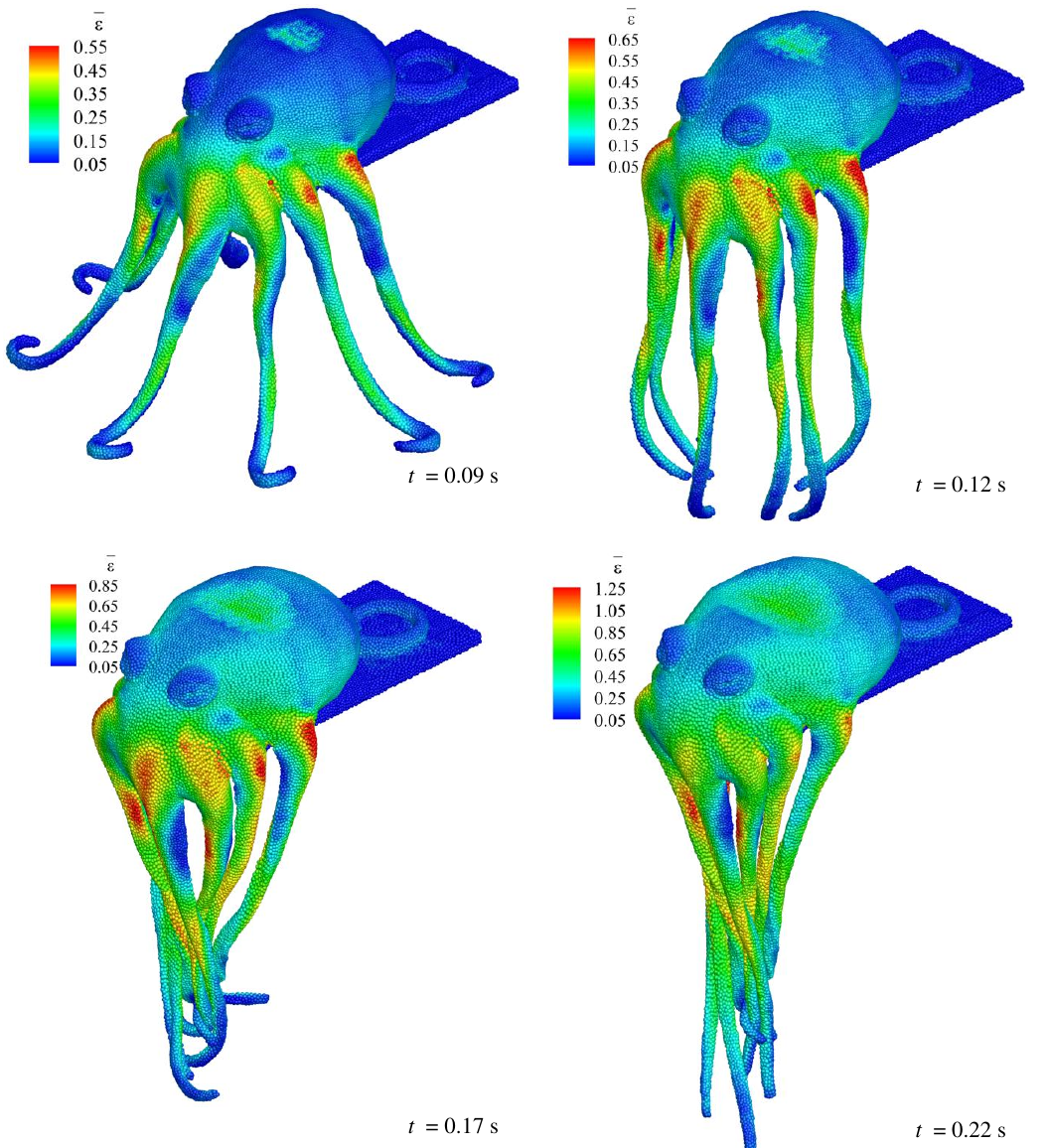}
	\caption{Oobleck octopus: 
		Deformed configuration of first stage colored by von Mises strain $\bar\epsilon$ 
		at serial temporal instants
		obtained by the present SPH-GENOG. }
	\label{figs:octopus_1st_stage}
\end{figure}
\begin{figure}[htb!]
	\centering
	\includegraphics[trim = 0mm 6mm 2mm 2mm, width=0.62\textwidth]{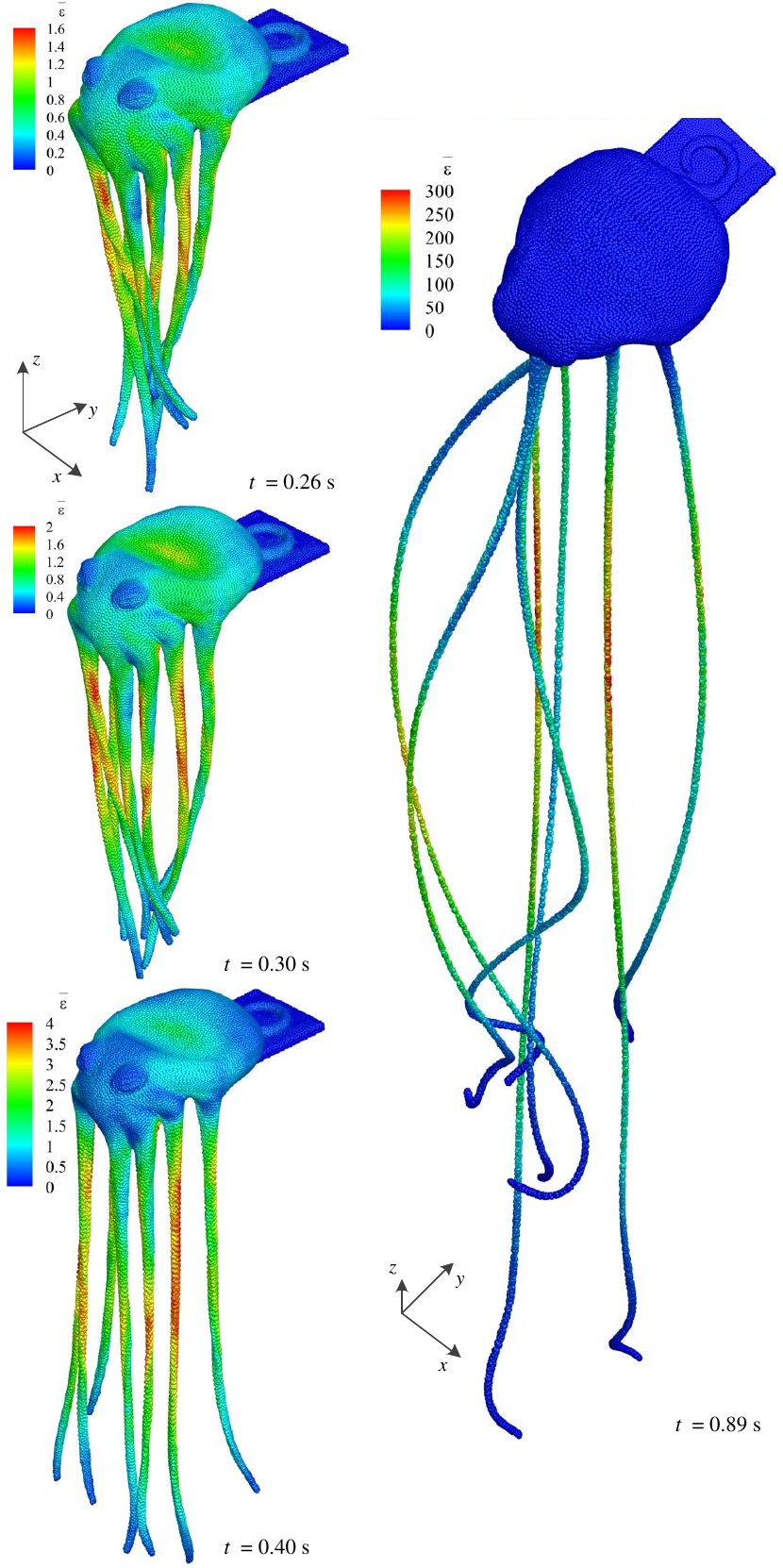}
	\caption{Oobleck octopus: Deformed configuration of second stage colored by von Mises strain $\bar\epsilon$ at serial temporal instants obtained by the present SPH-GENOG. }
	\label{figs:octopus_2nd_stage}
\end{figure}

Figure \ref{figs:octopus_1st_stage} shows 
the first stage of the octopus deformation, 
wherein the octopus feet collide with each other at high velocity. 
Moving on to the second stage, 
as illustrated in Fig. \ref{figs:octopus_2nd_stage}, 
significant plastic flow is observed in the octopus. 
The smooth deformation and huge strain fields 
highlight the potential of 
the present formulation for real-world applications.

\section{Concluding remarks}\label{sec:conclusion}

In conclusion, 
our presented essentially non-hourglass formulation 
provides a general solution in addressing the instability of hourglass modes 
for TLSPH solid dynamics with large deformations. 
By adopting a generalized approach 
based on volumetric-deviatoric stress decomposition, 
our formulation demonstrates versatility across a wide range of materials,
including elasticity, plasticity, and anisotropy. 
Through comprehensive validation using benchmark cases, 
together with a single set of modeling parameters, 
we establish the robustness and accuracy of the present formulation. 
Furthermore, 
the successful simulation of the very large deformation of Oobleck
serves as a compelling demonstration of 
the potential of the formulation in real-world scenarios. 
Note that the current formulation, 
initially designed for TLSPH, 
can be extended to accommodate the updated Lagrangian SPH,
making it suitable for situations 
where an updated Lagrangian approach is preferred.

%
%
\section*{CRediT authorship contribution statement}
{\bfseries  D. Wu:} Conceptualization, Methodology, Investigation, Visualization, Validation, Formal analysis, Writing - original draft, Writing - review and editing; 
{\bfseries  X.J. Tang:} Investigation, Writing - review and editing;
{\bfseries  S.H. Zhang:} Investigation, Writing - review and editing;
{\bfseries  X.Y. Hu:} Supervision, Methodology, Investigation, Writing - review and editing.
%
%
\section*{Declaration of competing interest }
The authors declare that they have no known competing financial interests 
or personal relationships that could have appeared to influence the work reported in this paper.
%
%
\section*{Acknowledgments}
D. Wu and X.Y. Hu would like to express their gratitude to the German Research Foundation (DFG) 
for their sponsorship of this research under grant number DFG HU1527/12-4.
\clearpage

\section*{References}
\vspace{-0.8cm}
\renewcommand{\refname}{}
\bibliographystyle{elsarticle-num}
\bibliography{IEEEabrv,mybibfile}

\newpage
\appendix
\section{Plastic algorithm}\label{app:plastic_algorithm}
While specific details regarding the non-linear hardening plastic model 
are available in our previous work \cite{tang2023explicit}
and insights into the viscous plastic model 
can be found in Ref. \cite{yue2015continuum}, 
we just focus on in-depth elaboration of the perfect 
and linear hardening plastic models in the following. 
Note that the primary distinctions among these four plastic models 
are specifically related to the return mapping 
which is employed to update the stress and strain states 
when a material undergoes deformation beyond its elastic limit.

The scalar yield function $f(\fancy{$\tau$}_{de})$ 
of the perfect and linear hardening plastic models
can be expressed as
\begin{equation}
	\label{yield_function}
	f (\fancy{$\tau$}_{de})
	=\|\fancy{$\tau$}_{de}\|_F
	- \sqrt{\frac{2}{3}} \left(\kappa \xi + \tau_y \right),
\end{equation}
where $\kappa$ is the hardening modulus, 
$\xi$ the hardening factor which is $0$ for perfect plasticity, 
and $\tau_y$ the initial flow stress, also called yield stress. 
Note that $ \|\bullet\|_{F} $ denotes a Frobenius norm of a tensor variable. 
The detailed algorithm of the linear hardening plastic model
from Ref. \cite{simo2006computational} 
is presented in Algorithm \ref{al:algorithm1}.
The superscript $(\bullet)^{trial}$ 
designates quantities pertaining to a trial elastic state
which is assessed to determine whether it exceeds the elastic limit, 
and the time stepping algorithm 
is performed in the elastoplastic material description. 
It is noteworthy that, 
since a position-based Verlet time-integration scheme \cite{zhang2021multi} 
is applied in this paper (see also Sec. \ref{sec:time_integration}), 
the plastic model algorithm is operated at the mid point of the $n$-th time step, 
i.e., the parameter is denoted by $(\bullet)^{n + \frac{1}{2}}$. 
{
	\linespread{1.0} \selectfont
	\begin{algorithm}
		\footnotesize
		Update deformation tensor $\mathbb{F}^{n+\frac{1}{2}}$
		
		Compute elastic predictor (Note that $\mathbb{C}_p^0 = \mathbb{I}$.)
		\begin{equation} 
			\fancy{$b$}_e^{trial, n+\frac{1}{2}}
			= \mathbb{F}^{n+\frac{1}{2}}  
			\left(\mathbb{C}_p^{n-\frac{1}{2}}\right)^{-1}
			\left(\mathbb{F}^{n+\frac{1}{2}}\right)^{\operatorname{T}},
			\nonumber
		\end{equation}
		\begin{equation} 
			\fancy{$\tau$}_{de}^{trial, n+\frac{1}{2}}
			= G \operatorname{dev} \left(
			\bar{\fancy{$b$}}_e^{trial, n+\frac{1}{2}}
			\right).
			\nonumber
		\end{equation}
		
		Check for plastic loading (Note that $\xi^0 = 0$.)
		\begin{equation}
			f^{trial, n+\frac{1}{2}}
			=\| \fancy{$\tau$}_{de}^{trial, n+\frac{1}{2}} \|_F
			- \sqrt{\frac{2}{3}} 
			\left(\kappa \xi^{n-\frac{1}{2}} + \tau_y \right).
			\nonumber
		\end{equation}
		
		\eIf{$ f^{trial, n+\frac{1}{2}} \leq 0 $} 
		{ Elastic state, 
			set $ (\bullet)^{n+\frac{1}{2}}
			= (\bullet)^{trial, n+\frac{1}{2}} $, 
			and $\left(\mathbb{C}_p^{n+\frac{1}{2}}\right)^{-1}
			=\left(\mathbb{C}_p^{n-\frac{1}{2}}\right)^{-1}$. }
		{Plastic state, and perform \ref{return_mapping} (the return mapping) }
		
		Compute normalized shear modulus \label{return_mapping}
		\begin{equation}
			\tilde{G}=\frac{1}{d}
			\operatorname{tr} \left(
			\bar{\fancy{$b$}}_e^{trial, n+\frac{1}{2}}
			\right) G.
			\nonumber
		\end{equation}
		Compute increment of hardening factor
		\begin{equation} 
			\xi^{incre, n+\frac{1}{2}}
			= \frac{0.5 f^{trial, n+\frac{1}{2}}}
			{\tilde{G} + \kappa/3.0}.
			\nonumber 
		\end{equation}
		Update hardening factor
		\begin{equation} 
			\xi^{n+\frac{1}{2}} 
			= \xi^{n-\frac{1}{2}} 
			+ \sqrt{\frac{2}{3}} \xi^{incre, n+\frac{1}{2}}.
			\nonumber 
		\end{equation}
		Update stress and deformation gradient
		\begin{equation} 
			\fancy{$\tau$}_{de}^{n+\frac{1}{2}} 
			=\fancy{$\tau$}_{de}^{trial, n+\frac{1}{2}}
			- 2 \tilde{G} \xi^{incre, n+\frac{1}{2}}
			\fancy{$\tau$}_{de}^{trial, n+\frac{1}{2}} 
			/ \| \fancy{$\tau$}_{de}^{trial, n+\frac{1}{2}} \|_F,
			\nonumber
		\end{equation}
		\begin{equation} 
			\fancy{$b$}_e^{n+\frac{1}{2}}
			= \frac{1}{G}\fancy{$\tau$}_{de}^{n+\frac{1}{2}}
			+ \frac{1}{d}
			\operatorname{tr} \left(
			\fancy{$b$}_e^{trial, n+\frac{1}{2}}
			\right) \mathbb{I},
			\nonumber
		\end{equation}
		\begin{equation}
			\left(\mathbb{C}_p^{n+\frac{1}{2}}\right)^{-1}
			= \left(\mathbb{F}^{n+\frac{1}{2}}\right)^{-1}
			\fancy{$b$}_e^{n+\frac{1}{2}}
			\left(\mathbb{F}^{n+\frac{1}{2}}\right)^{-\operatorname{T}}.
			\nonumber
		\end{equation}
		\caption{Algorithm for $ J_2 $ flow theory with linear isotropic hardening.
		}\label{al:algorithm1} 
	\end{algorithm}
}
%
\end{document}